# Exploration and Exploitation in Consumer Automation:

# Visualizing IoT Interactions with Topological Data Analysis

Thomas P. Novak
The George Washington University

Donna L. Hoffman
The George Washington University

This article proposes a method to uncover opportunities for exploitation and exploration from consumer IoT interaction data. We develop a unique decomposition of cosine similarity that quantifies exploitation through functional similarity of interactions, exploration through cross-capacity similarity of counterfactual interactions, and differentiation of the two opportunities through within-similarity. We propose a topological data analysis method that incorporates these components of similarity and provides for their visualization. Functionally similar automations reveal exploitation opportunities for substitutes-in-use or complements-in-use, while exploration opportunities extend functionality into new use cases. This data-driven approach provides marketers with a powerful capability to discover possibilities for refining existing automation features while exploring new innovations. More generally, our approach can aid marketing efforts to balance these strategic opportunities in high technology contexts.

Thomas P. Novak (novak@gwu.edu) is Denit Trust Distinguished Scholar and Professor of Marketing at The George Washington University School of Business, Washington, DC 20052. Phone: 202-994-8042

Donna Hoffman (dlhoffman@gwu.edu) is Louis Rosenfeld Distinguished Scholar and Professor of Marketing at The George Washington University School of Business, Washington, DC 20052. Phone: 202-994-3137

Acknowledgements. The Ayasdi software platform for topological data analysis (ayasdi.com) was used to construct the topological model of the IFTTT data. The authors acknowledge the support of Gunnar Carlsson, Co-Founder and President, Ayasdi, Inc., Menlo Park, Ca. IFTTT published applet data from 2011-2016 were provided by, and used with permission of Linden Tibbets, Co-Founder and CEO, IFTTT.com, San Francisco, CA.

# Exploration and Exploitation in Consumer Automation:
# Visualizing IoT Interactions with Topological Data Analysis

## ABSTRACT

This article proposes a method to uncover opportunities for exploitation and exploration from consumer IoT interaction data. We develop a unique decomposition of cosine similarity that quantifies exploitation through functional similarity of interactions, exploration through cross-capacity similarity of counterfactual interactions, and differentiation of the two opportunities through within-similarity. We propose a topological data analysis method that incorporates these components of similarity and provides for their visualization. Functionally similar automations reveal exploitation opportunities for substitutes-in-use or complements-in-use, while exploration opportunities extend functionality into new use cases. This data-driven approach provides marketers with a powerful capability to discover possibilities for refining existing automation features while exploring new innovations. More generally, our approach can aid marketing efforts to balance these strategic opportunities in high technology contexts.

**INTRODUCTION AND MOTIVATION**

The Internet of Things (IoT) offers significant opportunities for automation and consumers are automating increasing aspects of their lives through repeated interactions with smart objects. It is crucial for marketers to identify current trends and latent opportunities that can keep pace with evolving customer needs in the face of rapidly evolving technologies (Wertenbroch 2019). What is not possible today may very likely be possible tomorrow. A conceptually anchored methodology that can analyze consumer IoT interaction data holds the potential for unlocking important marketing insights.

We approach IoT interactions using assemblage theory (Delanda 2011, Hoffman and Novak 2018). In assemblage theory, interaction occurs through the paired directional capacities of the parts of an assemblage, with an assemblage emerging from the interactions of its parts. In the consumer IoT, common assemblages consist of two parts, an "if-part" and a "then-part". For example, the if-part, "if the dishwasher is finished running," is paired with the then-part, "then send me a text message". Not only does the dishwasher (if-part) have the capacity to affect the messaging app (then-part), but the messaging app also has the capacity to be affected by the dishwasher.

Upon reading this example, the reader might observe that dishwashers and messaging apps are not typically designed to interact in this manner. For this example to represent an assemblage, an additional part is needed, such as the Web service IFTTT. IFTTT is a popular IoT automation service that allows consumers to build if-then rules that enable interactions among hundreds of different devices and services, regardless of whether they were originally designed to interact with each other (Tibbets 2018). IFTTT works with over 800 apps and services and has been used by over 25 million consumers to automate various aspects of their lives (IFTTT 2023). Using IFTTT, a consumer can code an if-then rule so that *if* the dishwasher is finished, *then* a text message is sent. In addition to allowing consumers to code if-then rules, IFTTT also enables the paired capacities required for the rules to actually work. Through IFTTT, the dishwasher and messaging app can interact, even if they were not natively designed to do so. Because of this enabling capacity, an IFTTT if-then rule defines an assemblage formed by the interaction among its parts. Thus, this article uses the term *IoT interaction* to refer to an if-then rule that defines a particular automation assemblage.

*Exploitation and Exploration in the IoT*

The IoT interactions that consumers create serve as the basis for marketers to understand how consumers are using the IoT. In this article, we characterize the IoT interactions that consumers create as either exploitative or exploratory behaviors. After creating an if-then rule that enables their dishwasher to message them when it is finished, the consumer may follow an exploitative approach and develop similar rules, so their washing machine and coffee pot also message them when finished. Alternatively, the consumer may take an exploratory approach and reassemble the if- and then-parts of the IoT interaction in new ways. By reversing the causal sequence in their existing rules, the consumer may develop a new way of controlling their home appliances by messaging the appliances.

As another example of exploitation, a firm may identify that in addition to using temperature to control their smart thermostats, consumers were also using humidity and air quality indices from weather apps to control their thermostats. This may lead the firm to develop a new multi-metric thermostat that responds not just to temperature, but also to humidity and air quality. Considering exploration, a smart car might automatically check-in its location when parked at a train station. Hypothetically reversing causal roles, the train station might affect the smart car by syncing the car's parking time with the train schedule ("if train is delayed, then extend parking meter time"). This hypothetical interaction can form the basis of a new smart service offering.

Operationalizing exploitation and exploration from consumer IoT interactions aids marketing efforts to balance these strategies in high technology contexts. Exploitation allows existing automation features to be refined, while exploration reveals innovative opportunities. While the idea of exploitation vs. exploration is well known in the management literature, the concepts are understudied in the marketing literature. Our approach provides a bridge between the two literatures. In general, the management literature discusses exploitative innovation that refines existing knowledge versus exploratory innovation that tries new approaches (March 1991). By analyzing IoT interactions according to this framework, we are able to compare and contrast consumers' device automations that follow more exploitative patterns (e.g. connecting new devices to existing ones) with more exploratory patterns (e.g. trying new device integrations).

*Methodological Approach*

Our methodological approach begins with quantifying the if- and then-parts of an IoT interaction using word embeddings. This allows us to determine the similarity between different interactions based on the similarity of their if- and then-parts. We show that similarity among IoT interactions can be decomposed into three separate components of functional, cross-capacity, and within similarity. Combining this decomposition of similarity with concepts from assemblage theory (Novak and Hoffman 2023), we use TDA Mapper (Singh, Memoli, and Carlsson 2007; Carlsson 2014; Ray and Trovati 2017) to construct a data-based topological model of IoT interactions that can be used to visualize opportunities for exploitation and exploration.

*Contribution*

This article provides a number of contributions. First, we show that similarity can be more readily interpreted when it is decomposed into three separate components. We show that all three components of similarity have a direct and meaningful interpretation as exploitation versus exploration opportunities, which allows macro-level strategy concepts in the management literature to be connected to micro-level consumer behavior in the marketing literature. Second, we introduce topological data analysis (TDA) to consumer behavior research as a novel methodology for representing the underlying structure of similarity data, and for providing a clear visualization of exploitation versus exploration. Lastly, we focus on understanding opportunities for exploitation and exploration that can be directly accessed for marketing and product development action. While our data consist of consumer IoT interactions, these contributions apply to many additional consumer contexts.

We organize the article as follows. In the next section we review the relevant literatures on automation, exploitation and exploration, and compositional similarity. Then we present our methodology and results. We conclude with a discussion of the implications of our results for research and practice.

# LITERATURE REVIEW

*Automation*

The decision making literature has largely focused on consumers' reactive and preferential responses to increasing automation. Notable in the aversion stream, Leung, Paolacci, and Puntoni (2018) have shown that consumers react against automation when identity motives are closely tied to consumption and Granulo, Fuchs, and Puntoni (2020) have demonstrated that consumers' preferences for human (vs automated) labor is tied to symbolic consumption contexts, owing to stronger uniqueness motives in those contexts. Relatedly, Longoni, Bonezzi, and Morewedge (2019) find that consumers react against medical AI because of a belief that AI is unable to take consumers' unique situations into account. Contributing to the appreciation stream, Castelo, Bos, and Lehmann (2019) find that when the task is objective, then automation is preferred to humans, as it is when numeric algorithmic advice is offered (Logg, Minson, and Moore (2019). Following a different paradigm, Novak and Hoffman (2023) used a data discovery approach to uncover four higher-order categories of automation practices that emerge from consumer IoT interactions.

We note that the above prior work largely focuses on consumer preferences for and adoption of automation. The current research extends this body of work by focusing on what happens once automation has been adopted. We define concepts of exploitation and exploration opportunities based on the individual behavior of consumers who are actively using IoT automation. This allows us to connect consumer-level ideas from the marketing literature with the strategic implications of exploitation and exploration as discussed in the management literature.

*Exploitation and Exploration*

March (1991) first introduced the organizational learning constructs of exploitation and exploration as crucial to organizations because firms have a tendency to emphasize exploitation strategies at the expense of exploration strategies. Exploitation refers to the process of utilizing existing resources, knowledge, and capabilities to improve efficiency and effectiveness within an organization. It typically involves refining, extending, and improving current products, services, and processes (Levinthal and March, 1993). In marketing, exploitation processes tend to be focused on activities such as optimizing marketing mix elements, enhancing customer relationships, and improving distribution channels (Sirmon et al., 2008). Exploration entails the pursuit of new knowledge, opportunities, and capabilities that may lead to future growth and competitive advantage. It involves experimentation, discovery, risk-taking, and innovation (March, 1991). Marketing exploration activities include identifying new market segments, developing novel products or services, and adopting innovative marketing strategies and tactics (Sirmon et al., 2008).

This multidisciplinary research area tends to be strategically focused, emphasizing the need to manage the balance between an organization's exploitation and exploration activities for optimal organizational effectiveness. The common thread connecting these multiple literatures is that exploitation refers to the exercise of known capacities or capacities that are already in use, while exploration signifies capacities that are new or novel. Along these lines, several studies have investigated the relationship between exploitation, exploration, and marketing performance. For example, Atuahene-Gima (2005) found that marketing managers who engage in both exploration and exploitation activities tend to achieve higher levels of product innovation and overall marketing performance. Similarly, Kyriakopoulos and Moorman (2004) and Zhang, Wu, and Cui (2015) demonstrated that organizations with a balanced approach to exploration and exploitation are more likely to succeed in new product development. Vila, Bharadwaj, and Bahadir (2015) argued that the challenges presented by emerging markets make it vital that marketing managers pursue exploitation and exploration strategies in concert. Studying emerging markets in Africa, Asia, Europe, and South America, these authors find that advertising and promotion exploitation strategies can expand the access of existing products while new product introduction exploration strategies can enhance product innovation, providing a mechanism for competition in difficult and uncertain environments. Yalcinkaya, Calantone, and Griffith (2007) find that a firm's capacities for exploitation lay the groundwork for exploration, which subsequently have a positive impact on innovation and performance.

IoT interaction data offer a unique opportunity to quantify strategic notions of exploitation and exploration at the level of individual interactions based on actual consumer behavior. This opportunity becomes apparent when one considers that conceptualizations of exploitation and exploration implicitly incorporate concepts of distance. For example, in the marketing literature, Kim and Atuahene-Gima (2010) define exploitation as leveraging current market information related to existing (i.e., similar) practices. On the other hand, exploration requires "very new and radical" (p. 522) information that extends beyond the boundaries of existing (i.e., dissimilar) products in the market.

The notion of distance is explicitly addressed in Li, Vanhaverbeke, and Schoenmaker's (2008) systematic review and synthesis of the extensive, discipline-spanning literature on exploitation and exploration. They define exploitation and exploration according to the "knowledge distance" (p. 116) between new and existing knowledge in the knowledge space.

They note that firms may locate knowledge by "local search," which produces knowledge similar to what is already known. In contrast, "distant search" yields knowledge dissimilar from the existing knowledge base. Exploitation and exploration can then be determined by how local (familiar) or distant (unfamiliar) is the search of the knowledge space, respectively. Critically for our approach, Li, et. al. (2008) stress that this conceptualization allows exploitation and exploration to be operationalized by degree rather than as an either/or dichotomy.

*Compositional Similarity*

Compositional similarity is the idea that similarity between two interactions can be broken down into meaningful components (Novak and Hoffman 2023). Compositional similarity has a long-standing history, with Tversky (1977) proposing a feature-based model in which an object's similarity is expressed as a linear combination of shared and unique features. Similarly, Shepherd and Arabie (1979) represented similarity between stimuli as the sum of weights of their common and distinct features. Tversky's original model has been expanded in various ways. Tversky and Gati (1982), for example, demonstrated that continuous dimensions like size and brightness could be translated into features and included in an additive similarity model. We also note that although Tversky's model is additive, Sjoberg (1975) reviewed a range of models of similarity based upon more complex functions that incorporate ratios. These approaches have in common the idea that similarity expresses fundamental relationships among two events.

Compositional similarity has been applied in various fields beyond psychology, including patents, activities, online advertising, and search (Broder et al., 2008; Nascimento et al., 2011; Verma and Varma 2011; Nurbakova et al., 2017). Distance metrics can be applied across different contexts, and many of these metrics allow for meaningful decomposition. For example,

City Block (L1) distance breaks down overall distance into the sum of distances across *k* dimensions, while Euclidean (L2) distance is a function of distances across *k* dimensions (square root of the sum of squares). Cosine similarity decomposes overall similarity of vector representations based on word embeddings into separate components (Novak and Hoffman, 2023). The psychological similarity of pairs of words (and phrases) is reflected in the cosine similarity of their corresponding word (and phrase) embeddings (Mikolov, Yih, and Zweig 2013; Levy and Goldberg 2014). This relationship supports our approach of meaningfully decomposing consumers' IoT interactions into parts and rendering them visually, allowing us to interpret them in the context of opportunities for exploitation and exploration.

## METHODOLOGY

### *Data and Embeddings*

We analyze 20,675 unique if-then rules created by IFTTT users from 2011 to 2016 (Novak and Hoffman 2023). The IFTTT dataset offers a diverse range of IoT interactions to examine exploitation and exploration opportunities. Our unit of analysis is an IoT interaction, and we will show how similarity of different IoT interactions can be understood in the context of opportunities for exploitation and exploration.

Word embeddings are high-dimensional vectors that form the building blocks for visual representations of interactions, where an interaction is defined by a set of words (Mikolov, et.al. 2013; Levy and Goldberg 2014). We use Novak and Hoffman's (2023) dataset of embedding representations of the if- and then-phrases of these interactions as our foundation (see Web

Appendix A for further details)[1]. For each of 20,675 if-then interactions we have a 25-dimensional numerical vector (embedding), **i**, representing the if-phrase of the interaction, and a 25-dimensional numerical vector, **t**, representing the then-phrase of the interaction. These if- and then-embeddings were constructed by averaging individual word embeddings within the if- and then-phrases. For example, the if-embedding for the phrase "if photo and video YouTube new public photo uploaded by you" averages the embeddings for the individual words "photo," video," "YouTube," "new", and so on, ignoring stop words like "and", "by", " if" and "then." Similarly, the then-embedding for the phrase "then notes Evernote create a link note" averages the word embeddings for "notes," "Evernote," "create," "link," and "note."

*Components of Similarity*

Novak and Hoffman (2023) demonstrate that the similarity of two IoT interactions, for simplicity notated "1" and "2" throughout, where each is represented by the sum of their if- and then-embeddings **i** + **t**, can be decomposed into three components: *functional similarity* (cos$_{FUNCTIONAL}$) of the two interactions, *cross-capacity similarity* (cos$_{CC}$) of the two interactions, and a function of each of the two interaction's own *within similarity*, $f(\text{cos}_{WITHIN})$. Equation 1 displays the decomposition:

---

[1] The if- and then- phrase text, embedding vectors, and Web Appendices can be downloaded from OSF: https://osf.io/tf4kr/

(1) $\text{cos}_{\text{WHOLE}} = \cos(\mathbf{i}_1 + \mathbf{t}_1, \mathbf{i}_2 + \mathbf{t}_2)$

$= (\text{cos}_{\text{FUNCTIONAL}} + \text{cos}_{\text{CC}}) / f(\text{cos}_{\text{WITHIN}})$

where:

$\text{cos}_{\text{FUNCTIONAL}} = (\cos(\mathbf{i}_1, \mathbf{i}_2) + \cos(\mathbf{t}_1, \mathbf{t}_2)) / 2$

$\text{cos}_{\text{CC}} = (\cos(\mathbf{i}_1, \mathbf{t}_2) + \cos(\mathbf{i}_2, \mathbf{t}_1)) / 2$

$\text{cos}_{\text{WITHIN}} = \cos(\mathbf{i}_k, \mathbf{t}_k)$ for a given interaction $k$, so for interactions k=1 and k=2:

$f(\text{cos}_{\text{WITHIN}}) = \sqrt{(1 + \cos(\mathbf{i}_1, \mathbf{t}_1) + \cos(\mathbf{i}_2, \mathbf{t}_2) + \cos(\mathbf{i}_1, \mathbf{t}_1) \cdot \cos(\mathbf{i}_2, \mathbf{t}_2))}$

While Novak and Hoffman (2023) identified these three components, they only analyzed functional similarity. They did not consider cross-capacity similarity to be "substantively meaningful," and stated that combining all three terms together was "complex and unclear." However, by linking similarity of consumer IoT interactions to the concepts of exploitation and exploration, all three components of similarity can be meaningfully clarified.

Specifically, we show below that exploitation can be operationalized by the functional similarity of two interactions, while exploration can be defined by their cross-capacity similarity. The sum of the two numerator terms in equation (1) is constrained by the denominator, and this constraint grows as the within similarity (internal homogeneity) of the two interactions approaches one. Additionally, as the within similarity nears one, the correlation between functional similarity and cross-capacity similarity also approaches one. These relationships, which follow from equation (1), help us understand how opportunities for exploitation and exploration increase as the if- and then-parts of interactions become more heterogeneous.

**Exploitation is Captured by Functional Similarity.** Functional similarity calculates the average similarity of the if-phrases ($\cos(\mathbf{i}_1, \mathbf{i}_2)$) and then-phrases ($\cos(\mathbf{t}_1, \mathbf{t}_2)$) of two IoT interactions. When two if-then interactions exhibit high functional similarity, they possess a strong semantic correspondence, as both their if-phrases and then-phrases have similar meanings. Functional similarity implies that the two interactions are performing analogous tasks.

Functionally similar interactions offer opportunities for exploitation. This component of similarity models the behavior of consumers seeking to deepen their range of IoT interactions. In exploitation, consumers will likely seek interactions that exercise if- and then-capacities in ways similar to their existing interactions. Opportunities for exploitation are increased when $cos_{FUNCTIONAL}$ between interactions is relatively high and the $cos_{CC}$ between interactions is relatively low. We illustrate this by example below.

**Exploration is Represented by Cross-Capacity Similarity**. Cross-capacity similarity, $cos_{CC}$, involves a hypothetical interaction that has not actually been exercised, but could potentially be exercised. Cross-capacity similarity is obtained as the functional similarity of a first interaction based on embeddings $i_1$ and $t_1$ with a hypothetical counterfactual version of a second interaction, in which the if- and then-roles of the second interaction are reversed ($i_2$ and $t_2$ are swapped to become $t_2$ and $i_2$). In other words, in the counterfactual version, the "then" causes the "if" instead of the original "if" causing the "then."

To clarify with a concrete example, imagine Interaction 1 where sending a text message (if) changes the color of lights (then), and Interaction 2 where sending a text message (if) turns off the lights (then). Interactions 1 and 2 have high functional similarity since the semantic meanings of their if- and then-phrases are very similar. But their cross-capacity similarity, as defined by equation (1), is very low. Now consider a hypothetical variation of Interaction 2, called Interaction 2*. In Interaction 2*, turning off lights (if) triggers sending a text message (then), effectively reversing the if- and then-roles. The functional similarity of Interactions 1 and 2* is now very low, but their cross-capacity similarity is very high.

Two interactions that have high cross-capacity similarity suggest opportunities for exploration. The counterfactual implied in one of the interactions reverses the if-then causal direction of the original interaction, which broadens the scope of possibilities. Other *existing* interactions with high functional similarity to hypothetical Interaction 2* may also have a high cross-capacity similarity to Interaction 1 and represent opportunities for exploration. Our definition of exploration employs the existing parts of an interaction and repurposes them in a particular way, by disassembling the assemblage and reassembling it in the opposite causal direction.

**The Potential for Exploitation Versus Exploration is Constrained by Within Similarity.** For any interaction, we can calculate the within similarity, $\cos_{WITHIN}$, between the if- and then-phrases of the interaction. This term is incorporated in the denominator of equation (1), $f(\cos_{WITHIN})$, which is the square root of one plus the sum of the within similarities of two different interactions and their product. Since $\cos_{WITHIN}$ for 99.98% of the 20,675 interactions is greater than or equal to zero, we assume that $\cos_{WITHIN}$ ranges between 0 and 1. Then, the denominator's minimum value is 1 when both interactions have zero within similarity and its maximum value is 2 when both interactions have a within similarity of one. The numerator is the sum of $\cos_{FUNCTIONAL}$ and $\cos_{CC}$, which as cosine similarities can vary individually between -1 and 1. However, since $\cos_{WHOLE}$ must be between -1 and 1, values of $f(\cos_{WITHIN})$ less than two necessarily constrain the range of values that $\cos_{FUNCTIONAL}$ and $\cos_{CC}$ can take, which has implications for opportunities for exploitation and exploration.

When $f(\text{cos}_{\text{WITHIN}})$ is at its minimum value of 1, both interactions are as internally heterogeneous as possible, meaning that the if- and then-phrases are semantically distinct and the within similarities of both interactions are zero. In that case, since $\text{cos}_{\text{WHOLE}}$ must range between -1 and 1, the sum of $\text{cos}_{\text{FUNCTIONAL}}$ and $\text{cos}_{\text{CC}}$ must also range between -1 and 1. In other words, either functional similarity or cross-capacity similarity can be high, but not both. The more internally heterogeneous the two interactions are, the more opportunities there are to differentiate opportunities for exploitation (i.e., high functional similarity) from exploration (i.e., high cross-capacity similarity).

When $f(\text{cos}_{\text{WITHIN}})$ is at its maximum value of 2, both interactions are as internally homogeneous as possible, meaning that the if- and then-phrases are semantically identical and the within similarities of both interactions are one. In that case, both $\text{cos}_{\text{FUNCTIONAL}}$ and $\text{cos}_{\text{CC}}$ can achieve the maximum value of 1 and neither are constrained. However, when the within similarity of both interactions is one, it can be shown that functional similarity is exactly equal to cross-capacity similarity. (See Web Appendix A for further details and derivations regarding effects of $\text{cos}_{\text{WITHIN}}$). It follows that as within similarity of both interactions approaches one, then both functional and cross-capacity interactions will have the same semantic meaning. In that case, opportunities for exploitation cannot be differentiated from opportunities for exploration.

## *Topological Model of Functional Similarity*

We utilized Ayasdi's implementation (Ayasdi 2018) of the Topological Data Analysis (TDA) Mapper algorithm (Singh, Memoli, and Carlsson 2007; Carlsson 2014; Ray and Trovati 2017) to visualize the underlying structure of 20,675 IoT interactions. Our goal was to uncover potential opportunities for exploitation and exploration (for a detailed explanation of our TDA methodology, please refer to Web Appendix B). Although various visualization methods could be employed, such as low-dimensional manifold mapping techniques like UMAP (Novak and Hoffman 2023) or traditional dimensionality reduction approaches like Principal Component Analysis (PCA) and Multidimensional Scaling (MDS), TDA Mapper provides specific advantages in depicting exploitation and exploration opportunities.

TDA Mapper generates a simplicial complex, a graph-like structure that allows for easy visualization and comprehension of the network structure underlying the connections among all interactions. This method enables the clear identification of contiguous sections in the topological model containing exploitation opportunities, as opposed to discontiguous sections of the network containing exploration opportunities. In contrast, linear (e.g., PCA) and nonlinear (e.g., UMAP) dimensionality-reduction techniques position data points in a low-dimensional space. But they do not reveal the network structure underlying these connections that would identify contiguous or discontiguous sections.

**RESULTS**

*Topological Model*

Figure 1 depicts the topological model, which is a condensed representation of our data consisting of 20,675 rows (IoT interactions) and 50 columns (25-dimensional if-phrase embeddings **i** and 25-dimensional then-phrase embeddings **t**). The TDA Mapper algorithm organizes these 20,675 interactions into a smaller set of 741 nodes based on their functional similarity. Each node contains multiple interactions and interactions can belong to more than one node. Nodes are connected if they share interactions, which forms a network that is plotted using a force-directed graph (e.g., Fruchterman and Reingold 1991). The network exhibits a clear structure, uncovering well-defined and interconnected groups of IoT interactions organized into clusters, loops, and several flares (Carlsson 2014).

--- Insert Figure 1 ---

The model highlights specific types of interactions (e.g., "if monitoring event, smart home action") and broad groupings (e.g., "Smart Home Automation"). Since the topological model is based on functional similarity, interactions located close together present opportunities for exploitation. For example, consumers involved with smart home automation would be likely to exploit the range of interactions at the button left portion of the topological model, while interactions involving relational aspects of the smart home would be exploited in the bottom right portion.

The nodes in the topological model are colored by the average within similarity of the interactions belonging to that node. Red nodes contain homogeneous interactions which on average have a high within similarity. Opportunities for exploitation versus exploration should be less distinct for such interactions. Blue nodes, on the other hand, contain heterogeneous interactions which on average have a low within similarity. Opportunities for exploitation versus exploration should be more readily identified for interactions in these nodes.

*Examples of Exploration and Exploitation*

The topological model shown in Figure 1, based on functional similarity, uncovers not only opportunities for exploitation but also possibilities for exploration. Consider a focal interaction situated in one of the nodes of the topological model. Exploitation opportunities will consist of nodes containing interactions with high functional similarity to the focal interaction. These nodes will likely be found in proximal sections of the network close to the focal interaction and located in contiguous nodes connected to each other and the node containing the focal interaction.

On the other hand, opportunities for exploration will consist of nodes containing interactions with high cross-capacity similarity to the focal interaction. As mentioned earlier, these are counterfactual versions of existing interactions that have high functional similarity to the focal interaction. These counterfactual versions innovatively apply existing if-then capacities by reversing their causal relationships. Nodes containing interactions with high-cross capacity similarity will generally be in distal sections of the network farther from the focal interaction and located in discontiguous nodes connected to each other but not to the node containing the focal interaction.

*Exploitation and Exploration in Action*

**Four Example Focal Interactions.** To illustrate how our approach can be applied, we selected four interactions situated in the bottom right section of the topological model as focal interactions. In these focal interactions, consumers receive real-time notifications when a smart device or service detects an event. We chose focal interactions with a wide range of within similarity of their if- and then-embeddings. Within similarity of the four focal interactions ranged from highly homogeneous (.972) to extremely heterogeneous (.088).

Four scatterplots, displayed in the top left of Figures 2a-2d, depict the functional similarity and cross-capacity similarity of each of the four focal interactions with all other interactions. Interactions below the main diagonal represent exploitation opportunities, while those above the main diagonal indicate exploration opportunities. A large circle denotes the focal interaction. We note that the focal interaction's functional similarity with itself is always unity, and its cross-capacity similarity with itself equals its within similarity.

--- Insert Figures 2a-2d ---

To the right of each scatterplot is a table with examples of interactions representing opportunities for exploitation and exploration. These interactions are actual if-then rules created by IFTTT users. For each focal interaction, these opportunities comprise the interactions with the highest functional (exploitation) and cross-capacity (exploration) similarities among all 20,675 interactions. Due to space limitations, we list only the first few to demonstrate the approach.

The bottom left panel of Figures 2a-2d highlights exploitation opportunities by coloring the topological model from Figure 1 according to the functional similarity of the focal interaction with all other interactions (shaded blue if $\cos_{FUNCTIONAL} > 0.7$). The bottom right panel displays exploration opportunities by coloring the topological model based on the cross-capacity similarity of the focal interaction with all other interactions (shaded green if $\cos_{CC} > 0.6$). A large black circle indicates the location of the focal interaction.

**Correlation of Functional and Cross-Capacity Similarity.** A distinct progression is evident across the four panels of scatterplots in Figure 2. In Figure 2a, the high within similarity (0.972) of the focal interaction results in a correlation of 0.976 between cos$_{\text{FUNCTIONAL}}$ and cos$_{\text{CC}}$, with exploitation and exploration opportunities being difficult to distinguish. A very high within similarity makes it impossible to reassemble an interaction in a different way. This means that exploration through reassembly is not possible, and exploration (cos$_{\text{CC}}$) is the same as exploitation (cos$_{\text{FUNCTIONAL}}$). Logically speaking, we only have opportunities for exploitation.

In Figure 2b, the lower within similarity (0.446) leads to a reduced correlation (0.574), resulting in interactions that either exhibit high functional similarity (below the diagonal) or high cross-capacity similarity (above the diagonal). This offers opportunities for both exploitation and exploration. Figure 2c displays an even lower within similarity (0.399) and a correlation of 0.045 between cos$_{\text{FUNCTIONAL}}$ and cos$_{\text{CC}}$, further emphasizing the separation of exploitation and exploration opportunities above and below the diagonal. Lastly, in Figure 2d, the within similarity (0.088) approaches zero, the correlation between cos$_{\text{FUNCTIONAL}}$ and cos$_{\text{CC}}$ is negative (-0.503), and the distinction between exploitation and exploration opportunities becomes even more clearly defined.

Collectively, the scatterplots in Figures 2a-d offer further insight into how within similarity constrains the potential for exploitation versus exploration. In Figure 2a, the within similarity of the focal interaction is almost one (0.972). While the within similarity of the 20,674 second interactions range from -0.010 to 1, with a mean of 0.509, the functional and cross-capacity similarity are nearly the same across 20,674 pairs of interactions (correlation = 0.976). This indicates that a single, very homogeneous car-to-car interaction ("if bmw speeding, send car notification") significantly constrains the ability to distinguish exploitation from exploration opportunities. Higher within similarity blurs the distinction between exploitation and exploration, with the blur taking literal form as a cloud of points tightly concentrated along the main diagonal. In contrast, the scatterplots in Figures 2b-d show that as within similarity decreases, our ability to differentiate exploitation from exploration opportunities increases dramatically as points move off the main diagonal. In Figure 2d, where within similarity (0.088) approaches zero, the effect of the constraint on the numerator of equation (1) is clearly visible in the scatterplot, where the points are either high on the horizontal or vertical dimension, but not both.

**Topological Models**. The topological models at the bottom of the four panels in Figure 2 show the locations of opportunities for exploitation (left) and exploration (right). In Figure 2a, consistent with the high 0.972 within similarity of the focal interaction, exploitation and exploration both occur in similar locations, particularly in the bottom right portion of the network that is contiguous with the focal interaction. To some extent, exploration also takes place in discontiguous sections of the network, primarily the bottom left section containing smart home automation applets. However, exploration still remains within the realm of smart home interactions and as noted earlier, is essentially the same as exploitation. By "contiguous," we mean a set of interconnected blue or green nodes that also connect to the focal interaction. By "discontiguous," we mean a set of interconnected blue or green nodes that do not connect to the focal interaction. As within similarity decreases from 0.445 to 0.088, the topological models at the bottom of Figures 2b through 2d tell a different story. For all three of these focal interactions, exploitation most often occurs in proximal and contiguous portions of the network, while exploration takes place in increasingly distal and discontiguous portions of the network.

The figures, particularly Figures 2b-d, where exploitation and exploration diverge, showcase examples that extend beyond the literal reversal of if- and then-phrases we mentioned earlier (e.g., an interaction where a text message triggers a change in lights becomes an interaction where changing lights prompt a text message to be sent). For instance, in Figure 2b, the focal interaction "if parrot flower power temperature alert, call phone" has a cross-capacity similarity of 0.726 with "if tomorrow's forecast calls for, snooze rainmachine."

*Generalizing Effects of Within Similarity on Exploitation Versus Exploration*

**Plotting the Similarity Relationship.** The sequence of plots in Figure 2a-d demonstrates that as the within similarity of the focal interaction approaches zero across four focal interactions, the correlation between functional and cross-capacity similarity becomes negative. This negative correlation in Figure 2d allows for a clear differentiation of opportunities for exploitation from those for exploration. However, what does the relationship between within similarity and the correlation of functional and cross-capacity similarity look like across all 20,675 interactions, not just in these four examples? Figure 3 displays this relationship. In Figure 3, each of the 20,675 interactions is considered in turn as the focal interaction. The within similarity of each focal interaction (horizontal axis) is plotted against the correlation of cos$_{\text{FUNCTIONAL}}$ and cos$_{\text{CC}}$ (vertical axis) for the focal interaction with each of the other 20,764 interactions. Figure 3 also highlights the four focal interactions from Figures 2a-d.

--- Insert Figure 3 ---

At relatively high levels of a focal interaction's within similarity, above .60, functional and cross-capacity similarity are highly correlated, generally well above .75. For these homogeneous focal interactions, exploitation is favored over exploration. At relatively low levels of a focal interaction's within similarity, for example below .40, the correlation between functional and cross-capacity similarity ranges between -.75 and .75. For these heterogeneous interactions, both exploitation and exploration exist. However, Figure 3 shows that even when within similarity is below the mean (which is .509 in these 20,675 interactions), there is still a range of variation in the correlation of functional and cross-capacity similarity for a given value of within similarity. Thus, the potential for distinguishing opportunities for exploitation from those for exploration tends to increase as within similarity approaches zero.

**Multiple Regression of the Similarity Relationship**. A series of nested regression models were used to investigate the relationship between the similarity of the parts within an interaction ($\cos_{WITHIN}$) and the correlation between that interaction's functional and cross-capacity similarity with other interactions ($r_{FUNCTIONAL,\ CC}$). The models suggest that as the within similarity of an interaction decreases, especially when $\cos_{WITHIN}$ is below the average, the ability to find opportunities for both exploitation and exploration increases. Furthermore, the regression models indicate that the ability to find opportunities for exploration also varies depending on the type of interaction, with type of interaction reflected by where the containing the interaction is located in the topological model. The effects of interaction type are most impactful when within similarity is below average. Overall, the regression analyses imply that lower within similarity and differences in interaction types allow for greater opportunities to exploit existing capabilities and explore new capacities. Web Appendix C contains full details of this analysis.

**DISCUSSION AND CONCLUSION**

In this article we develop an approach to quantify concepts of exploitation and exploration from IoT interaction data. Our unique and highly interpretable decomposition of the components of cosine similarity allows us to locate automations that represent opportunities for exploitation (functional similarity) and exploration (cross-capacity similarity), as limited by the similarity of an interaction's parts (within similarity). We show that by conditioning on focal interactions, we can systematically proceed from the known automations consumers are constructing to additional similar automations that both exploit current uses, as well as those potential automations that explore what is possible.

*Implications*

Analysis of functional similarity can inform firm efforts to improve device interfaces, offer more customized automation options based on usage patterns, or upsell compatible devices. Exploitation based on functional similarity takes advantage of what is already known from the data to refine and improve current offerings in existing IoT markets. Exploration, on the other hand, uncovers novel combinations of interactions that suggest new ways consumers can potentially interact with devices. Cross-capacity similarity reveals new combinations of consumer behaviors that are distant from consumers' existing IoT interactions. The marketing value of cross-capacity interactions lies not in the relatively trivial reversal of if- and then-phrases, but in using the hypothetical scenario to identify actual reassembled interactions that are conceptually related and ripe for exploration. These potential interaction combinations can provide inspiration for opportunities to, for example, bundle devices in new ways or build new platforms for connecting devices.

Functionally similar interactions representing opportunities for exploitation can serve as either substitutes-in-use or complements-in-use (Shocker, Bayus, and Kim 2004). Consider the focal interaction from Figure 2b ("if parrot flower power temperature alert, call phone"). Some opportunities for exploitation are *functional complements* whose "if" and "then" services are closely related to the focal interaction, but exercise complementary capacities (e.g., "if parrot flower power soil moisture alert, call phone"). A consumer creating the focal interaction might create the second interaction as a complement-in-use, with both delivering device notifications by calling the phone. Other IoT interactions are *functional substitutes* that exploit the focal interaction's capacities for notification, but in a different way (e.g., "if parrot flower power temperature alert, send sms"). Here, the user transitions from phone calls to sms notifications. Opportunities for exploration, on the other hand, complement existing uses to *extend functionality* into new use cases (e.g., "if phone call answered, turn on wemo power switch"). Now the phone call serves as a trigger for another action. Our approach can shed insight into how IoT interactions are substitutes or complements. An important research question is what compositional models of cosine similarity best differentiate substitutes from complements.

### *Extensions To Our Approach*

**Alternate Data.** Our approach for identifying opportunities for exploitation and exploration can be readily extended beyond simple if-then interactions. For example, IFTTT recently introduced new paid features for creating advanced automations, including the ability to create multiple "then" events. By using query and filter events, IFTTT users can also create multiple "if" events that produce complex conditional automations. Similarly, Apple's Shortcuts and the automation services Zapier also allow multiple if and then events. Newer approaches to automation based upon generative AI introduce even richer capacities. Auto-GPT (Significant-Gravitas 2023) chains together "thoughts" from a large language model (LLM) to autonomously achieve a goal. Similarly, AgentGPT (reworkd 2023) allows a user to configure and deploy autonomous AI agents that complete a goal that is broken down into a sequence of tasks. All of these examples are based upon text input that is chained together in various ways and generate interaction data amenable to a generalization of our methodology.

More broadly, our approach to exploration and exploitation can be applied to any consumer context in which the notion of reversing if and then roles makes sense. The marketing opportunities here are voluminous. In content discovery in music streaming services, if you frequently listen to a specific music genre, then similar genres are suggested. This can be logically reversed so that if you explore a new genre, this is treated as a frequent listen. The reversed interaction alters the consumer's profile to suggest a broader range of music interests, in turn impacting future recommendations in the direction of exploration. As another example, a consumer at a travel site may first book a flight to a destination, and then be offered suggestions for hotels at that destination. This process could be reversed so that consumers who book hotels are offered various flight options. The ability to obtain recommendations in both directions will facilitate both opportunities for exploitation and exploration in travel decisions.

**Methodological Directions.** We trained our word embeddings on a domain-specific text corpus of IoT interactions. Alternatively one could use transfer learning (e.g. Ruder et al 2019) to fine-tune pre-trained word embeddings on our IoT interaction data. Word embeddings obtained through transfer learning benefit from the larger and more diverse dataset the embeddings were originally trained on, capturing a broader range of semantic information. Word embeddings obtained through transfer learning would be expected to have a much higher dimensionality (300 dimensional vectors are typical) than our 25-dimensional vectors. However, the meaning of the words in our corpus is to some degree idiosyncratic to the context of it-then rules and may not be accurately captured by higher-dimensional vectors trained on a much broader text corpus. Nonetheless, future research could compare the relative advantages of lower-dimensional domain-specific embeddings versus high-dimensional embeddings obtained through transfer learning.

While our embeddings were learned for individual words in our corpus and then averaged within if- and then- phrases, other researchers might want to consider approaches like SBERT (Reimers and Gurevych 2019) that train embeddings at the phrase level. Future research might also explore whether vectors of probabilities from topic models (e.g., Tirunillai and Tellis 2014) or keyword-assisted topic models (Eshima et al 2023) could be used as an alternative to embeddings. Topic models are best viewed as an alternative to rather than a substitute for embeddings, since as Novak and Hoffman (2023) noted, embeddings and topic models can produce different results. Given the inherent interpretability of topic models, it may be useful to compare topic probabilities to embeddings as the basis for a topological model.

Last, we call for future research to compare human judgements with machine predictions of exploration and exploitation. Such research could validate the extent to which the construct of functional similarity corresponds to human judgments of exploration, and cross-capacity similarity corresponds to human judgments of exploitation. Additionally, future research can verify that specific interactions predicted to be opportunities for exploration (exploitation) based on high cross-capacity (functional) similarity are perceived by consumers to be opportunities for exploration (exploitation).

**Concluding Thoughts.** IoT interaction data derived directly from consumer behaviors contributes to strategic efforts to find the right balance between exploitation and exploration in high technology contexts. By using such data to refine and optimize existing automation features (exploitation), companies can improve the efficiency and effectiveness of their products and services. At the same time, IoT interaction data can reveal opportunities for new automation capabilities and innovations (exploration), allowing companies to adapt to changing customer needs and stay ahead of competitors. This balance between exploitation and exploration can lead to the development of more adaptive and intelligent automation systems that evolve alongside consumer needs and technological advancements.

# REFERENCES

Atuahene-Gima, Kwaku (2005), "Resolving the Capability-Rigidity Paradox in New Product Innovation," *Journal of Marketing*, 69(4), 61-83.

Ayasdi (2018), "Ayasdi Machine Intelligence Platform, V7.12.0," Ayasdi, Inc. Menlo Park, CA

Broder, Andrei, Peter Ciccolo, Evgeniy Gabrilovich, and Bo Pang (2008), "Domain-Specific Query Augmentation Using Folksonomy Tags: The Case of Contextual Advertising," In *Proceedings of the 1st Workshop on Information Retrieval for Advertising*.

Carlsson, Gunnar, (2014), "Topological Pattern Recognition for Point Cloud Data," *Acta Numerica,* 23, 289-368.

Castelo, Noah., Bos, Maarten W., & Lehmann, Donald R (2019), "Task-Dependent Algorithm Aversion," *Journal of Marketing Research*, *56*(5), 809–825.

DeLanda, Manuel (2016), *Assemblage Theory*, Edinburgh, UK: Edinburgh University Press.

Eshima, Shusei, Kosuke Imai, and Tomoya Sasaki (2023), "Keyword-Assisted Topic Models," *American Journal of Political Science*, February, 1-21, https://doi.org/10.1111/ajps.12779

Fruchterman, Thomas M. J. and Reingold, Edward M. (1991), "Graph Drawing by Force-Directed Placement," *Software: Practice and Experience*, 21(11), 1129-1164.

Granulo, Armin, Christoph Fuchs, and Stefano Puntoni (2021), "Preference for Human (vs. Robotic) Labor is Stronger in Symbolic Consumption Contexts," *Journal of Consumer Psychology*, 31(1), January, 72-80.

Hoffman, Donna L., and Thomas P. Novak (2018), "Consumer and Object Experience in the Internet of Things: An Assemblage Theory Approach," *Journal of Consumer Research,* 44(6), 1178-1204.

IFTTT (2023), https://ifttt.com/explore/best-no-code-app-builder, accessed September 13, 2023.

Kim, Namwoon and Kwaku Atuahene-Gima (2010), "Using Exploratory and Exploitative Market Learning for New Product Development," *Journal of Product Innovation Management*, 27(4), July, 519-536.

Kyriakopoulos, Kyriakos and Christine Moorman (2004), "Tradeoffs in Marketing Exploitation and Exploration Strategies: The Overlooked Role of Market Orientation," *International Journal of Research in Marketing*, 21(3), September, 219-240.

Leung, Eugina, Gabriele Paolacci, and Stefano Puntoni (2018), "Man Versus Machine: Resisting Automation in Identity-Based Consumer Behavior," *Journal of Marketing Research*, 55(6), 818-831.

Levinthal, Daniel A. and March, James G. (1993), "The Myopia of Learning," *Strategic Management Journal*, 14(S2), 95-112.

Levy, Omer and Yoav Goldberg (2014), "Linguistic Regularities in Sparse and Explicit Word Representations," *Proceedings of the Eighteenth Conference on Computational Natural Language Learning*, Baltimore, MD, Association for Computational Linguistics, 171-180.

Li, Ying, Wim Vanhaverbeke, and Wilfred Schoenmaker (2008), "Exploration and Exploitation in Innovation: Reframing the Interpretation," *Creativity and Innovation Management*, 17(2), June, 95-168.

Logg, Jennifer M., Julia A. Minson, and Don A. Moore (2019), "Algorithm Appreciation: People Prefer Algorithmic to Human Judgment," *Organizational Behavior and Human Decision Processes,* 151 (March), 90-103.

Longoni, Chiara, Andrea Bonezzi, and Carey K. Morewedge (2019), "Resistance to Medical Artificial Intelligence," *Journal of Consumer Research,* 46(4), 629-650.

March, James G. (1991), "Exploration and Exploitation in Organizational Learning," *Organization Science*, 2(1), February, 71-87.

Mikolov, Tomas, Ilya Sutskever, Kai Chen, Gregory S. Corrado, and Jeffrey Dean, (2013), "Distributed Representations of Words and Phrases and Their Compositionality," Advances in Neural Information Processing Systems 26: 27th Annual Conference on Neural Information Processing Systems 2013. Proceedings of a meeting held December 5-8, 2013, Lake Tahoe, Nevada, United States , pages 3111–3119.

Mikolov, Tomas, Wen-tau Yih, and Geoffrey Zweig (2013), "Linguistic Regularities in Continuous Space Word Representations," *Proceedings of the 2013 Conference of the North American Chapter of the Association for Computational Linguistics: Human Language Technologies*, pages 746–751, Atlanta, Georgia, June. Association for Computational Linguistics.

Nascimento, Cristiano, Laender, Alberto H. F., da Silva, Altigran S., and Gonçalves, Marcos André (2011), "A Source Independent Framework for Research Paper Recommendation," In *Proceedings of the 11th Annual International ACM/IEEE Joint Conference on Digital Libraries*, June, 297-306.

Novak, Thomas P., and Donna L. Hoffman (2023), "Automation Assemblages in the Internet of Things: Discovering Qualitative Practices at the Boundaries of Quantitative Change," *Journal of Consumer Research,* 49 (5), 811-837.

Nurbakova, Diana, Laporte, Léa, Calabretto, Sylvie, and Gensel, Jerome (2017), "Recommendation of Short-Term Activity Sequences During Distributed Events," *Procedia Computer Science*, 108, 2069-2078. 10.1016/j.procs.2017.05.154.

Ray, Jeffrey and Marcello Trovati (2017), "A Survey of Topological Data Analysis (TDA) Methods Implemented in Python," in *International Conference on Intelligent Networking and Collaborative Systems*, 594-600. Springer, Cham.

reworkd (2023), "AgentGPT," Accessed on May 2, 2023. https://agentgpt.reworkd.ai/

Reimers, Nils, and Iryna Gurevych (2019), "Sentence-BERT: Sentence Embeddings Using Siamese BERT-Networks," arXiv preprint arXiv:1908.10084

Ruder, Sebastian, Matthew E. Peters, Swabha Swayamdipta, and Thomas Wolf (2019), "Transfer Learning in Natural Language Processing" In Proceedings of the 2019 conference of the North American chapter of the association for computational linguistics: Tutorials, pp. 15-18.

Shepard, Roger N., and Phipps Arabie (1979), "Additive Clustering: Representation of Similarities as Combinations of Discrete Overlapping Properties," *Psychological Review* 86 (2), 87-123.

Shocker, Allan D., Barry L. Bayus, and Namwoon Kim (2004), "Product Complements and Substitutes in the Real World: The Relevance of 'Other Products'," *Journal of Marketing* 68(1), 28-40.

Significant-Gravitas (2023), "Auto-GPT: An Autonomous GPT-4 Experiment." Accessed May 2, 2023. https://github.com/Significant-Gravitas/Auto-GPT

Singh, Gurjeet, Facundo Mémoli, and Gunnar E. Carlsson. (2007), "Topological Methods for the Analysis of High Dimensional Data Sets and 3d Object Recognition," In *SPBG*, pp. 91-100. Soulier et. al 2012

Sirmon, David G., Hitt, Michael A., Ireland, R. Duane, & Gilbert, Brett Anitra (2008), "Resource Orchestration to Create Competitive Advantage: Breadth, Depth, and Life Cycle Effects," *Journal of Management*, 34 (5), 886-907.

Sjoberg, Lennart. "Models of Similarity and Intensity," *Psychological Bulletin,* 82 (2), 191-206.

Tibbets, Linden (2018), "Connecting Everything with Everything: The Sky is the Limit," *GfK Marketing Intelligence Review*, 11(2), November, 48-53.

Tirunillai, Seshadri, and Gerard J. Tellis (2014), "Mining Marketing Meaning from Online Chatter: Strategic Brand Analysis of Big Data Using Latent Dirichlet Allocation," *Journal of Marketing Research*, 51 (4), 463-479.

Tversky, Amos (1977), "Features of Similarity," *Psychological Review*, 84(4), 327.

Tversky, Amos, and Itamar Gati (1982), "Similarity, Separability, and the Triangle Inequality." *Psychological Review* 89 (2), 123-154.

Verma, Manisha, and Vasudeva Varma (2011), "Patent Search Using IPC Classification Vectors," In *Proceedings of the 4th Workshop on Patent Information Retrieval*, ACM, 9-12.

Vila, Omar Rodriquez, Sundar G. Bharadwaj, and S. Cem Bahadir (2015), "Exploration- and Exploitation-Oriented Marketing Strategies and Sales Growth in Emerging Markets," *Customer Needs and Solutions*, 2, 277-289.

Wertenbroch, Klaus (2019), *"*From the Editor: A Manifesto for Research on Automation in Marketing and Consumer Behavior," *Journal of Marketing Behavior, 4 (1), 1-10*.

Yalcinkaya, Goksel, Roger J. Calantone, and David A. Griffith (2007), "An Examination of Exploration and Exploitation Capabilities: Implications for Product Innovation and Marketing Performance," *Journal of International Marketing*, 15(4), 63-93.

Zhang, Haisu, Fang Wu, and Anna Shaojie Cui (2015), "Balancing Marketing Exploration and Market Exploitation in Product Innovation: A Contingency Perspective," *International Journal of Research in Marketing*, 32(3), September, 297-308.

**Figure 1**

**Topological Model of 20,675 IoT Interactions**

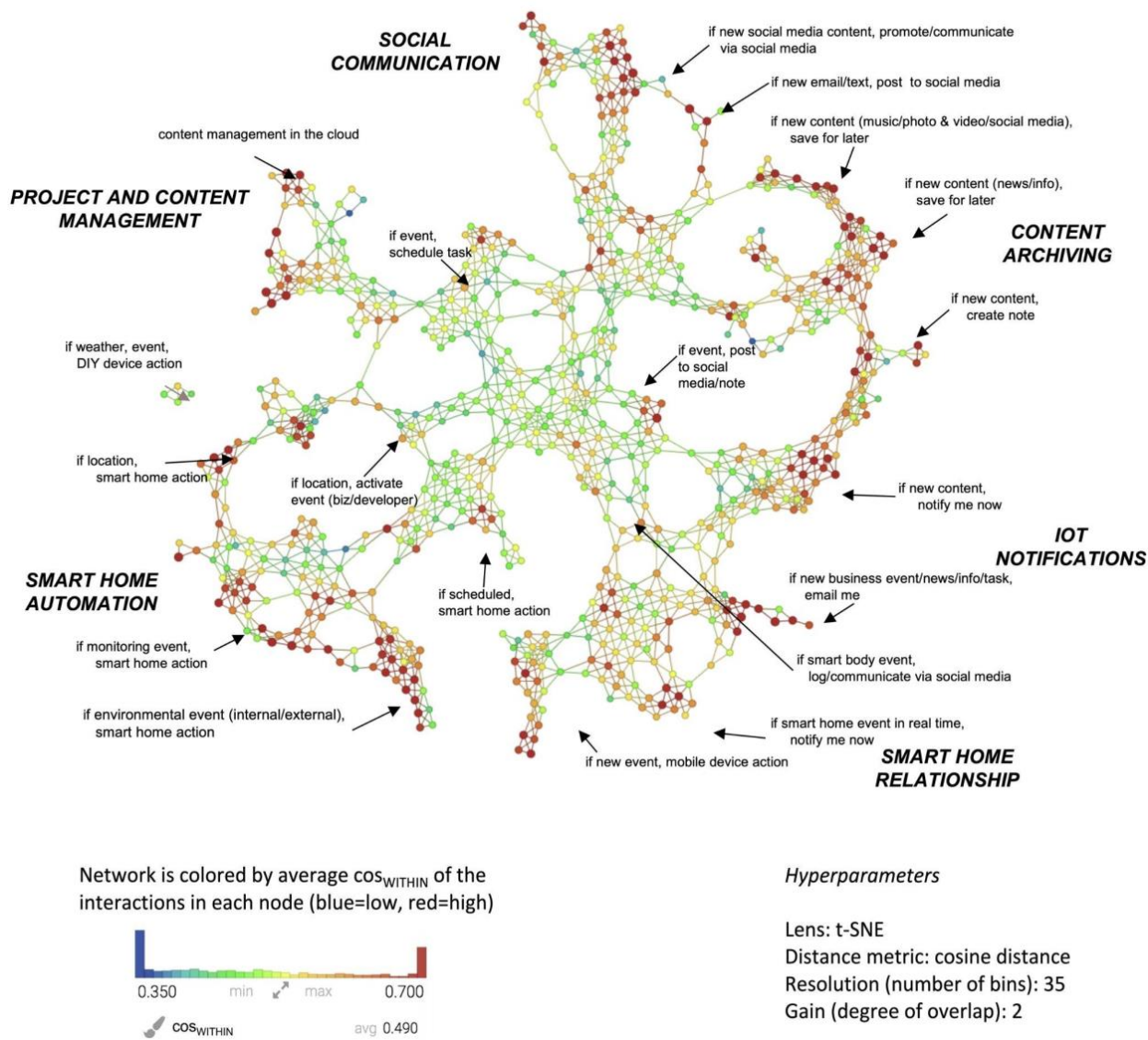

The topological model is a network of 741 overlapping nodes representing 20,675 IoT interactions. Nodes are connected if they have interactions in common. Singletons represented by 10 nodes incorporating 53 interactions are not shown.

## Figure 2a - Exploitation and Exploration Opportunities for Focal Interaction 17312: "if bmw speeding, send car notification"

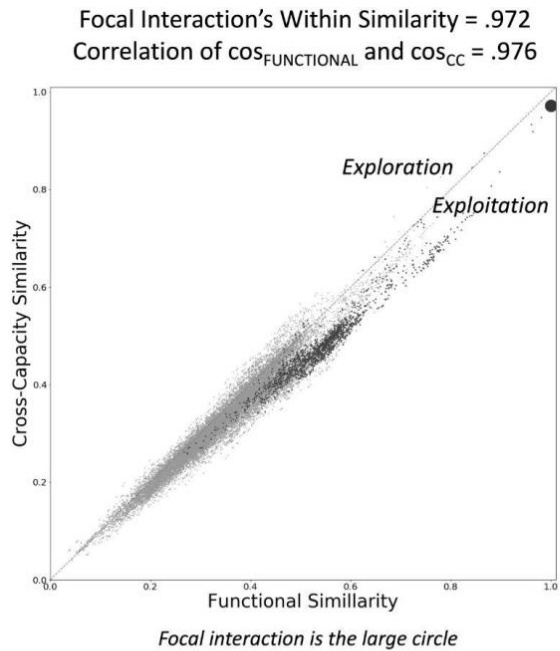

Focal Interaction's Within Similarity = .972
Correlation of $\cos_{FUNCTIONAL}$ and $\cos_{CC}$ = .976

Focal interaction is the large circle

| $\cos_{FUNCT}$ | $\cos_{CC}$ | Opportunities for Exploitation and Exploration |
|---|---|---|
| .981 | .947 | If bmw arrive soon, send car notification (2563) |
| .881 | .807 | If bmw enter area, send tv notification (5012) |
| .783 | .686 | if low fuel level send notification (6804) |
| .745 | .646 | If automatic check engine light turn on text me (3608) |
| .704 | .666 | If bmw enter area turn lifxZ light on (5019) |

**Exploitation Opportunities**
Network is colored by $\cos_{FUNCTIONAL}$ of Focal Interaction 17312 (large black circle) with all other Interactions.

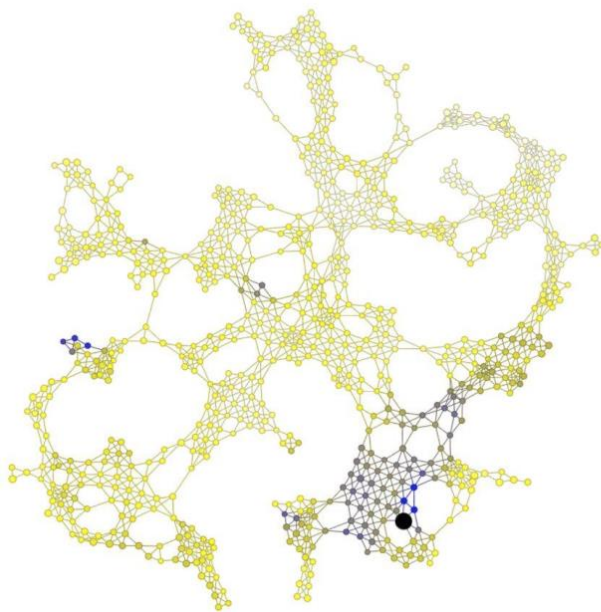

**Exploration Opportunities**
Network is colored by $\cos_{CC}$ of Focal Interaction 17312 (large black circle) with all other Interactions.

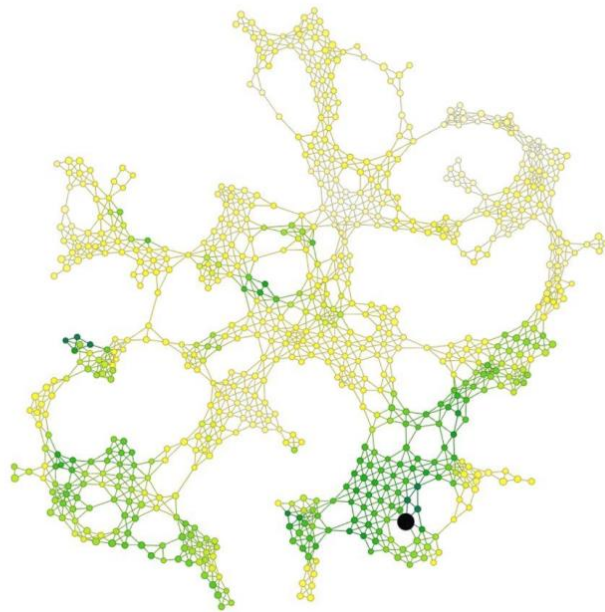

## Figure 2b - Exploitation and Exploration Opportunities for Focal Interaction 18053: "if parrot flower power temperature alert, call phone"

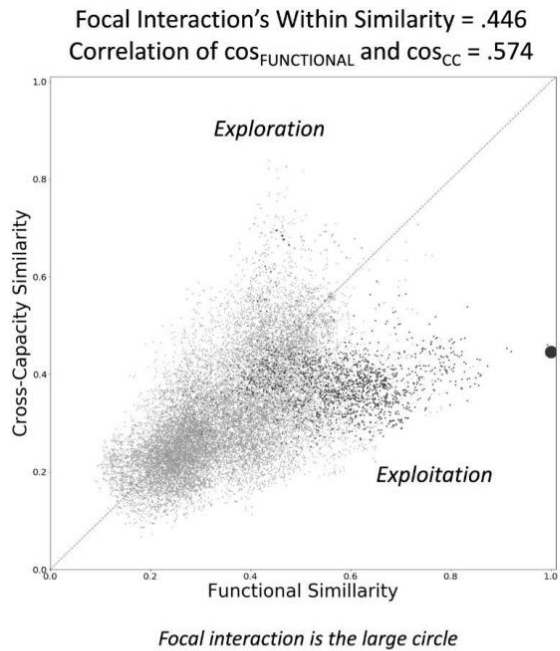

Focal Interaction's Within Similarity = .446
Correlation of $\cos_{FUNCTIONAL}$ and $\cos_{CC}$ = .574

| $\cos_{FUNCT}$ | $\cos_{CC}$ | Opportunities for Exploitation |
|---|---|---|
| .993 | .460 | If parrot flower power soil moisture alert, call phone (17216) |
| .874 | .417 | if parrot flower power temperature alert, send sms (18055) |
| .808 | .420 | If ge appliance cook preheat temperature achieved, call phone (15874) |

| $\cos_{FUNCT}$ | $\cos_{CC}$ | Opportunities for Exploration |
|---|---|---|
| .437 | .838 | If phone call answered, turn on wemo power switch (2370) |
| .471 | .824 | If phone call missed specific number, turn on dlink smart power plug (15646) |
| .511 | .726 | If tomorrows forecast call for, snooze rainmachine (18746) |

*Focal interaction is the large circle*

**Exploitation Opportunities**
Network is colored by $\cos_{FUNCTIONAL}$ of Focal Interaction 18053 (large black circle) with all other Interactions.

**Exploration Opportunities**
Network is colored by $\cos_{CC}$ of Focal Interaction 18053 (large black circle) with all other Interactions.

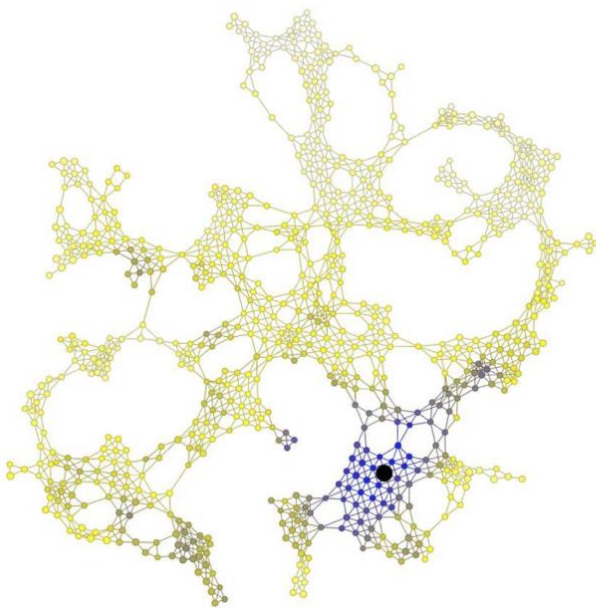
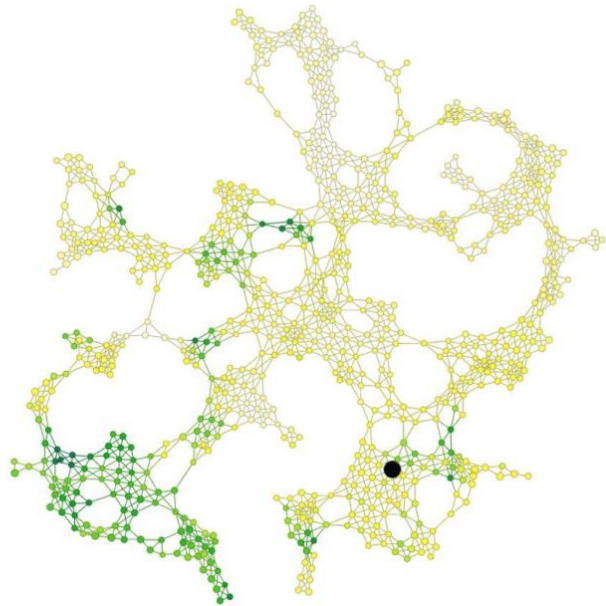

**Figure 2c - Exploitation and Exploration Opportunities for Focal Interaction 11162:**
*"if new weemo motion after quiet period, send email"*

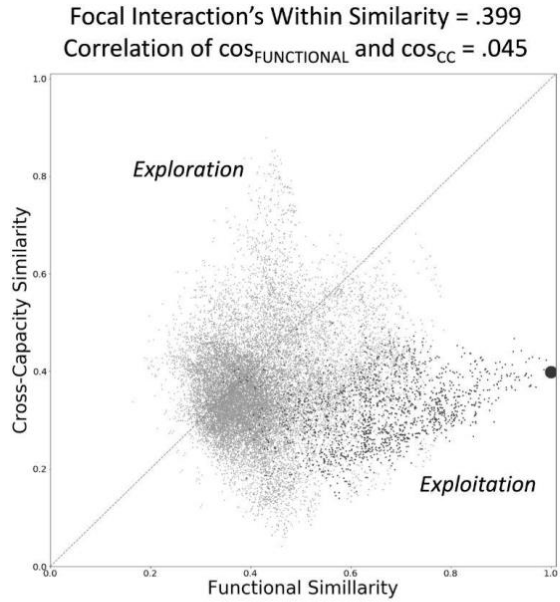

Focal Interaction's Within Similarity = .399
Correlation of $cos_{FUNCTIONAL}$ and $cos_{CC}$ = .045

| $cos_{FUNCT}$ | $cos_{CC}$ | Opportunities for Exploitation |
|---|---|---|
| .979 | .367 | If komfy switch camera motion detect, send email (7007) |
| .950 | .418 | If new ring detect, send email (7687) |
| .900 | .389 | if homeboy low battery, send email (6783) |

| $cos_{FUNCT}$ | $cos_{CC}$ | Opportunities for Exploration |
|---|---|---|
| .431 | .879 | if ifttt email, start record manything (16838) |
| .454 | .768 | If new email, switch on wattio power pod (8768) |
| .436 | .738 | If new email, turn on lightwaveRF light (867) |

*Focal interaction is the large circle*

**Exploitation Opportunities**
Network is colored by $cos_{FUNCTIONAL}$ of Focal Interaction 11612 (large black circle) with all other Interactions.

**Exploration Opportunities**
Network is colored by $cos_{CC}$ of Focal Interaction 11612 (large black circle) with all other Interactions.

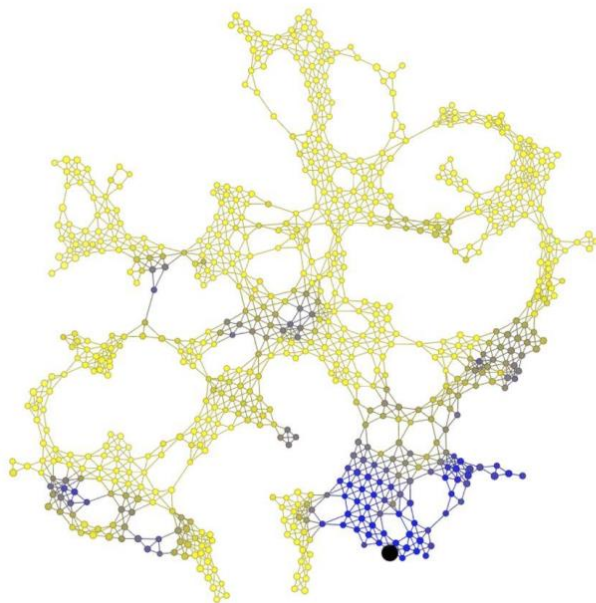
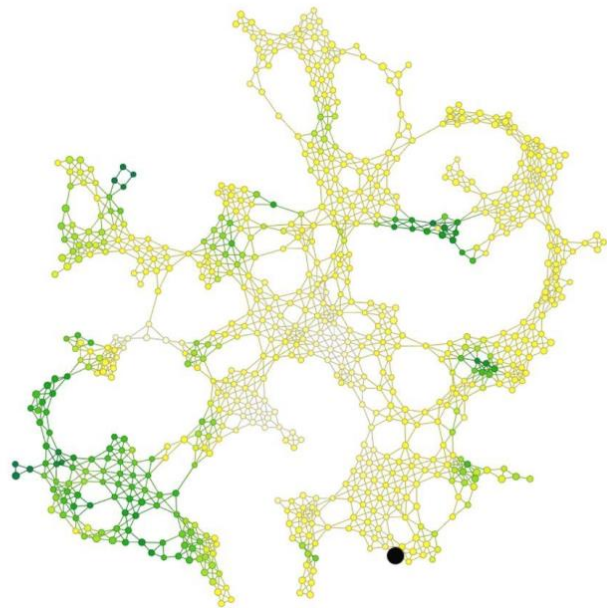

## Figure 2d - Exploitation and Exploration Opportunities for Focal Interaction 3968: "if current weather condition change, send gmail"

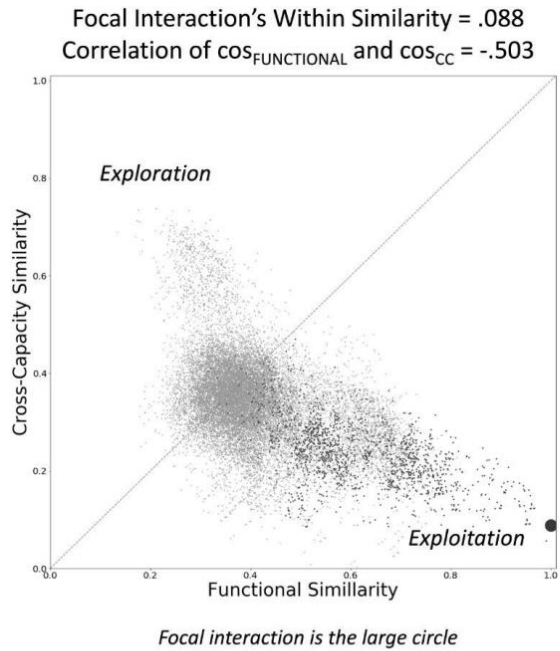

Focal Interaction's Within Similarity = .088
Correlation of $\cos_{FUNCTIONAL}$ and $\cos_{CC}$ = -.503

| $\cos_{FUNCT}$ | $\cos_{CC}$ | Opportunities for Exploitation |
|---|---|---|
| .965 | .123 | If sunrise send gmail (17459) |
| .922 | .168 | If netatmo weather station wind speed rises above threshold, send gmail (19371) |
| .893 | .135 | if netatmo weather station carbon dioxide rises above threshold, send email (3400) |

| $\cos_{FUNCT}$ | $\cos_{CC}$ | Opportunities for Exploration |
|---|---|---|
| .212 | .737 | If new gmail, set temperature nest thermostat (8760) |
| .245 | .733 | If new gmail, set ecobee comfort profile for n hour (8755) |
| .252 | .705 | if ifttt email, change color Philips hue lighting (16781) |

*Focal interaction is the large circle*

### Exploration Opportunities
Network is colored by $\cos_{CC}$ of Focal Interaction 3968 (large black circle) with all other Interactions.

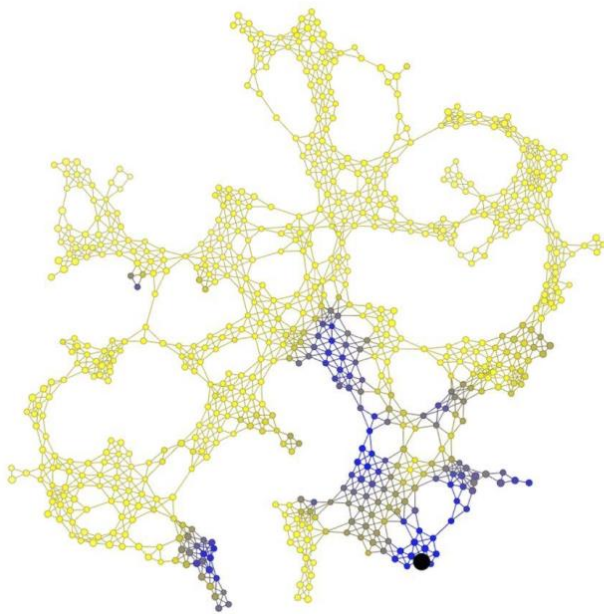

### Exploration Opportunities
Network is colored by $\cos_{CC}$ of Focal Interaction 3968 (large black circle) with all other Interactions.

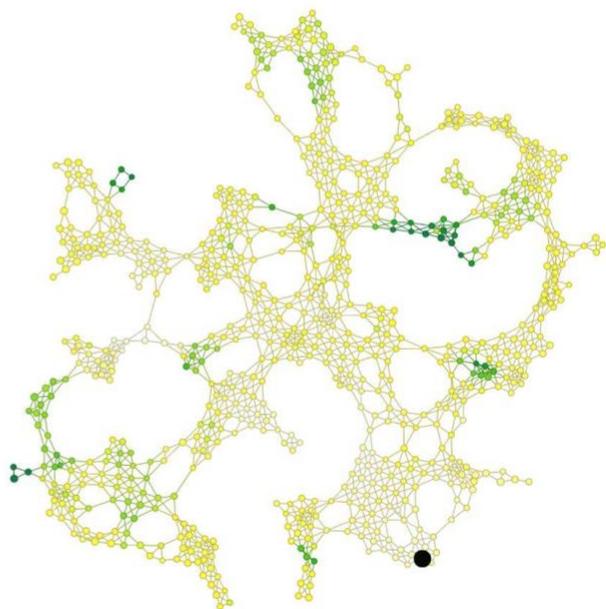

**Figure 3**

**Opportunities for Exploitation and Exploration for Each of 20,675 Interactions**

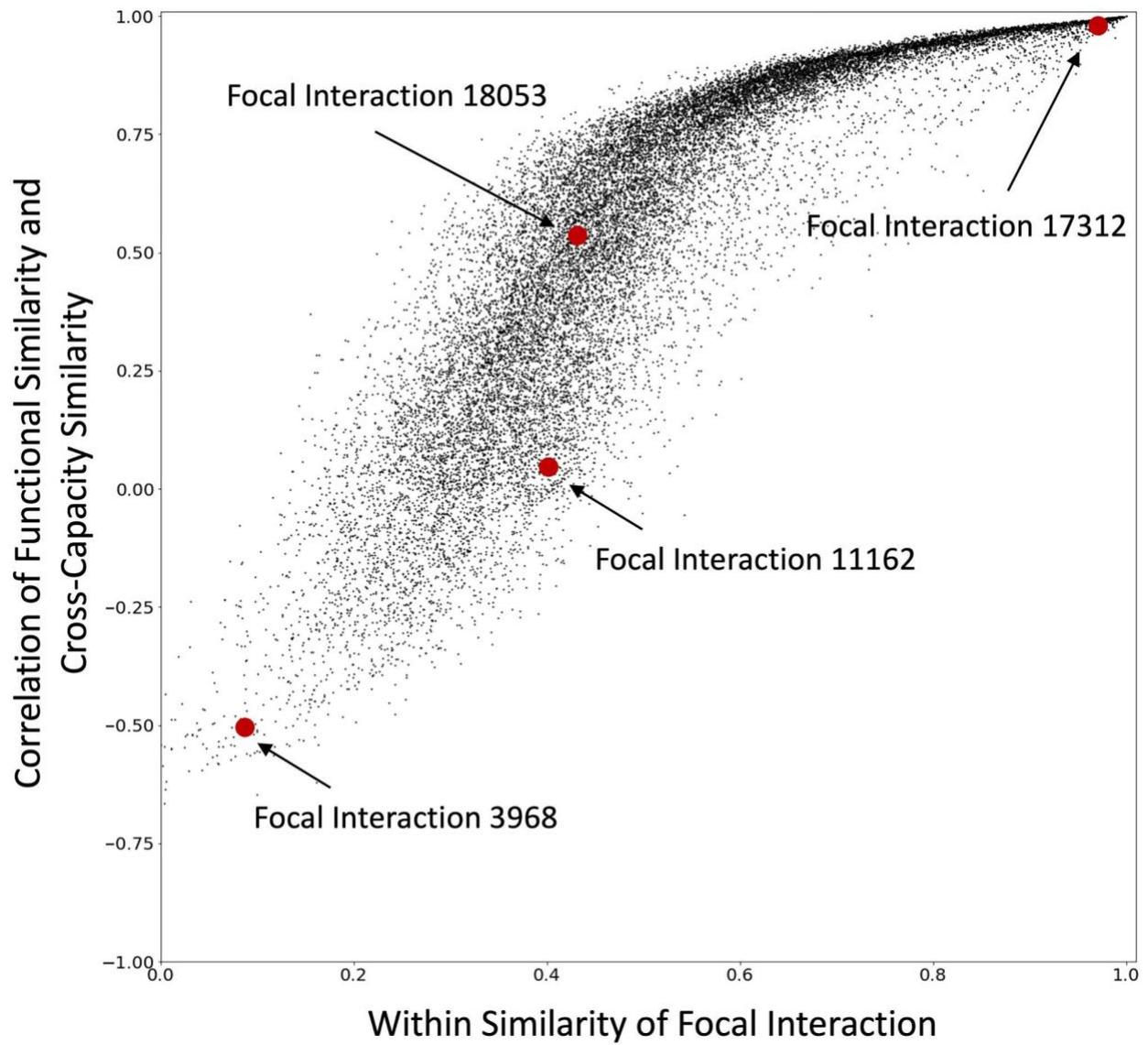

# Web Appendix for

# "Exploration and Exploitation in Consumer Automation: Visualizing IoT Interactions with Topological Data Analysis"

Thomas P. Novak

The George Washington University

Donna L. Hoffman

The George Washington University

**Web Appendices**

- Web Appendix A: Within Similarity
- Web Appendix B: TDA Mapper Methodology
- Web Appendix C: Multiple Regression Predicting Correlation of Functional and Cross-Capacity Similarity from Within Similarity

# Web Appendix A: Within Similarity

Equation (1) from the paper, showing the decomposition of overall similarity ($\cos_{WHOLE}$) into components, is reproduced below:

(A1) $\quad \cos_{WHOLE} = (\cos_{FUNCTIONAL} + \cos_{CC}) \,/\, f(\cos_{WITHIN})$

where:

$$f(\cos_{WITHIN}) = \sqrt{(1 + \cos(\mathbf{i}_1,\mathbf{t}_1) + \cos(\mathbf{i}_2,\mathbf{t}_2) + \cos(\mathbf{i}_1,\mathbf{t}_1)\cdot\cos(\mathbf{i}_2,\mathbf{t}_2))}$$

To illustrate how the denominator of equation (A1), $f(\cos_{WITHIN})$, affects the value of $\cos_{WHOLE}$ as two interactions vary in their within similarity, consider Table A1. Table A1 displays the values of $f(\cos_{WITHIN})$ for different combinations of within similarity for Interaction 1 (rows) and Interaction 2 (columns), where within similarity for each interaction is set to 0, .5 or 1. While within similarity can be negative (as low as -1), we note that in our data within similarity across 20,675 interactions ranged from -.01 to 1, with a mean of with a mean of 0.509. Since 99.98% of the 20,675 interactions had within similarity greater than or equal to zero, we restrict attention to non-negative values of within-similarity.

As the within similarity increases in either or both interactions, the value of $f(\cos_{WITHIN})$ also increases. This increase in the denominator, $f(\cos_{WITHIN})$, will result in a decrease in the value of $\cos_{WHOLE}$, assuming that the numerator ($\cos_{FUNCTIONAL} + \cos_{CC}$) remains constant. By examining Table A1, the impact of within similarity on $f(\cos_{WITHIN})$ and consequently on $\cos_{WHOLE}$ provides insights into the balance between exploration and exploitation based on the similarity of the interactions.

**Table A1.** Values of $f(\cos_{WITHIN})$ for Levels of Within Similarity

| Interaction 1 \ Interaction 2 | $\cos(i_2,t_2) = 0$ | $\cos(i_2,t_2) = 0.5$ | $\cos(i_2,t_2) = 1$ |
|---|---|---|---|
| $\cos(i_1,t_1) = 0$ | 1 | $\sqrt{1.5}$ | $\sqrt{2}$ |
| $\cos(i_1,t_1) = 0.5$ | $\sqrt{1.5}$ | $\sqrt{2.25}$ | $\sqrt{3}$ |
| $\cos(i_1,t_1) = 1$ | $\sqrt{2}$ | $\sqrt{3}$ | 2 |

First consider when the denominator, $f(\cos_{WITHIN})$, is at its minimum value of one. This happens when both interactions are as internally heterogeneous as possible (the if- and then-phrases are semantically distinct and the within similarities of both interactions are zero). In this case, since $\cos_{WHOLE}$ must range between -1 to 1, the sum of $\cos_{FUNCTIONAL}$ and $\cos_{CC}$ must also be within the same range. This means that when $f(\cos_{WITHIN})$ is at its minimum, the values of $\cos_{FUNCTIONAL}$ and cosCC must balance each other, so their sum remains between -1 and 1. Either functional similarity (exploitation opportunities) can be high, or cross-capacity similarity (exploration opportunities) can be high, but not both. The more heterogeneous the if- and then-phrases of the two interactions are the more opportunities to differentiate exploitation from interaction opportunities.

Second consider when the denominator, $f(\cos_{WITHIN})$, is at its maximum value of two. This happens when both interactions are as internally homogeneous as possible (the if- and then-phrases are semantically identical and the within similarities of both interactions are one). In this case both $\cos_{FUNCTIONAL}$ and $\cos_{CC}$ can achieve the maximum value of 1, since after dividing by two in the denominator, $\cos_{WHOLE}$ will be bounded by -1 and 1. It additionally is the case that when both interactions have a within-similarity of one, then functional similarity is equal to cross-capacity similarity.

The equality of functional and cross-capacity similarity when within similarity is one can be easily shown. First, we set:

$$\cos(\mathbf{i}_1, \mathbf{t}_1) = \cos(\mathbf{i}_2, \mathbf{t}_2) = 1$$

This implies that $\mathbf{i}_1$ and $\mathbf{t}_1$ are identical, and that $\mathbf{i}_2$ and $\mathbf{t}_2$ are identical. However, note that $\mathbf{i}_1$ can differ from $\mathbf{i}_2$, and $\mathbf{t}_1$ can differ from $\mathbf{t}_2$. Now, consider $\cos_{FUNCTIONAL}$ and $\cos_{CC}$:

$$\cos_{FUNCTIONAL} = (\cos(\mathbf{i}_1, \mathbf{i}_2) + \cos(\mathbf{t}_1, \mathbf{t}_2))/2$$

$$\cos_{CC} = (\cos(\mathbf{i}_1, \mathbf{t}_2) + \cos(\mathbf{i}_2, \mathbf{t}_1))/2$$

Since both $\mathbf{i}_1 = \mathbf{t}_1$ and $\mathbf{i}_2 = \mathbf{t}_2$, we can rewrite both $\cos_{FUNCTIONAL}$ and $\cos_{CC}$ as follows, demonstrating their equivalence:

$$\cos_{FUNCTIONAL} = (\cos(\mathbf{i}_1, \mathbf{i}_2) + \cos(\mathbf{i}_1, \mathbf{i}_2))/2 = \cos(\mathbf{i}_1, \mathbf{i}_2)$$

$$\cos_{CC} = (\cos(\mathbf{i}_1, \mathbf{i}_2) + \cos(\mathbf{i}_2, \mathbf{i}_1))/2 = \cos(\mathbf{i}_1, \mathbf{i}_2)$$

Thus, when the within similarity approaches one for both interactions, opportunities for exploitation cannot be differentiated from opportunities from exploration.

## Web Appendix B: TDA Mapper Methodology

TDA Mapper Implementation

We used the TDA Mapper algorithm (Singh, Memoli and Carlsson 2007; Ray and Trovati 2017) to visualize the 20,675 IoT interactions, based on their if- and then-embeddings as constructed by Novak and Hoffman (2023). Briefly, these embeddings, **i** and **t**, were constructed as the normed average of the word embeddings, **w**, of the $n_i$ words in the if-phrase of the interaction, and the $n_t$ words in the then-phrase of the interaction thus:

(B1a)  $\mathbf{i} = \mathbf{i}^* / \|\mathbf{i}^*\|$, where $\mathbf{i}^* = \sum \mathbf{w} / n_i$, so that $\|\mathbf{i}\| = 1$

(B1b)  $\mathbf{t} = \mathbf{t}^* / \|\mathbf{t}^*\|$, where $\mathbf{t}^* = \sum \mathbf{w} / n_t$, so that $\|\mathbf{t}\| = 1$

The word embeddings are 25-feature vectors, **w**, learned with word2vec (Mikolov, et.al. 2013a). Note that the if- and then-embeddings, **i** and **t**, as well as the individual word embeddings, **w**, all lie in the same underlying 25-dimensional feature space.

The superiority of the 25-dimensional solution for these data is documented in Web Appendix A of Novak and Hoffman (2023) (see: https://osf.io/vh48r/ ). Novak and Hoffman extracted 10, 25, 50, 100, 200 and 300 feature solutions for word embeddings. Based on the downstream task of how well a reduced two-dimensional UMAP solution predicts the proportion of times a word appears in the if- or then-phrase of an IFTTT rule, both Random Forest Regression and Support Vector Regression found the highest $R^2$ was obtained with 25-feature embeddings. While we used t-SNE coordinates, rather than UMAP coordinate, as the lenses for the TDA Mapper analysis, we do not anticipate appreciably different results for t-SNE vs. UMAP.

However, to be sure, we also compared Random Forest regressions with 10-fold cross validation using the two t-SNE coordinates to predict the criterion of the proportion of times each word appeared in the if- or then- phrase of the IFTTT rule. Results again supported 25-feature embeddings, for all values of t-SNE perplexity ranging from 5 to 50 (the value of perplexity used in our paper is 15).

We used Ayasdi's commercial implementation of TDA Mapper to construct a topological network (Ayasdi 2018), but a wide range of open source solutions are available[1]. The input data for TDA Mapper is a rectangular matrix of point cloud data. The rows of the rectangular matrix are the 20,675 interactions, and the columns are the 50 learned features from the if- and then-embeddings (25 if- and 25 then-features). Along with specifying the data, TDA Mapper requires that we specify a distance metric. We used cosine distance. Cosine distances for all pairs of interactions were obtained as one minus functional cosine similarity.

The first step of TDA Mapper uses two functions, or *lenses,* to map the 50 columns onto two numbers. While any two lenses can be used, a good lens will capture aspects of the point cloud data relevant to a particular research problem. Our problem involves representing the structure of the functional similarities among the 20,675 interactions, which is a 20,675 x 20,675 matrix of $\cos_{FUNCTIONAL}$ between each of the nearly 214 million pairs of interactions. Thus, as potential lenses, we considered six different machine learning methods for dimensionality reduction that are commonly used in practice: PCA, MDS, Locally Linear Embedding, Isomap, Spectral Embedding, and t-SNE (perplexity equal to 15). For each of the six methods, we used the first two coordinates of the 20,675 interactions as lenses.

In the second step, TDA Mapper sorts the 20,657 interactions into overlapping *bins*, based on the values of the two lenses. This was separately done for each of the six methods for dimensionality reduction. The process of sorting interactions into overlapping bins requires specifying hyperparameters for resolution and gain. We selected a resolution of 35 (providing an initial sorting of interactions into $35^2$ = 1025 overlapping bins based on the lenses) and a gain of 2 (allowing a 50% overlap of bins, where percent overlap = 1 - 1/gain).

---

[1] Open source Python code for implementing TDA mapper is available as Kepler Mapper (https://kepler-mapper.scikit-tda.org/en/latest/) which is a library in the scikit-tda project (https://scikit-tda.org/). Additional Python TDA tools include pyTDA (https://github.com/stephenhky/PyTDA), Python Mapper (http://danifold.net/mapper/), Knotter (https://github.com/rosinality/knotter), ReNom (https://github.com/ReNom-dev-team/ReNom), and Dionysus (https://pypi.org/project/dionysus/). R packages include the R interface for GUDHI, Dionysus, PHAT, and TDA mapper. In addition, the JavaPlex library implements persistent homology for MATLAB and java-based systems and CTL is a C++ library for computational topology.

In the third step, TDA Mapper clusters the interactions within each bin based on the original columns (i.e., the 25 if-embeddings and the 25 then-embeddings). The clustering is based on the chosen distance metric, which is functional cosine distance. Interactions within each of the 1025 overlapping bins defined by the lenses are clustered by their functional cosine distance using single-linkage clustering. While Ayasdi's implementation uses single-linkage, other implementations allow different clustering methods to be used at this stage. However, keep in mind that a relatively small number of interactions are being clustered within each of the 1025 bins.

The fourth step of TDA Mapper constructs a network representation as a topological model of the data. Each within-bin cluster of interactions defines a node in the topological network. Thus, nodes represent sets of interactions that are similar in terms of functional cosine distance. Since the data were divided into overlapping bins, interactions can be in multiple nodes. Identical nodes are merged and nodes with at least one interaction in common are joined by an edge. The resulting topological model is a network of overlapping clusters of interactions. The network is plotted using a force-directed graph to provide a visualization. A key advantage of TDA is that the shape of the topological model is guided by an image of the rows of the data (i.e., the interactions) as viewed through a lens that reduces the dimensionality of the column variables (i.e., the two coordinates from a particular dimensionality reduction technique), while at the same time incorporating the high-dimensional information in the original column variables (i.e., the 25 if-embeddings and the 25 then-embeddings).

Comparison of Topological Models Across the Six Lenses

Figure B1 plots the 20,675 interactions on two lenses (i.e., the first two coordinates) using each of the six machine learning methods for dimensionality reduction: PCA, MDS, Locally Linear Embedding, Isomap, Spectral Embedding, and t-SNE. Cosine distances for all pairs of interactions were obtained (see equation A1). All models were fit in Python using PCA and manifold learning algorithms from scikit-learn 0.19.1. Figure 2 plots the 20,675 IoT interactions for each of the six models. The interactions in Figure B1 are colored by the within similarity of each interaction's if- and then- embeddings, with blue indicating higher within similarity (i.e., homogeneous interactions), and orange indicating lower within similarity (i.e., heterogeneous interactions). We note that Figure B1 is the conventional way the similarities among these interactions would be visualized using various machine learning approaches for dimensionality reduction.

As seen in Figure B1, all six pairs of lenses broadly distinguish interactions in terms of the similarity of their if- and then- phrases. That is, each of the six plots in Figure B1 show two blue clouds separated by an orange cloud. However, none of the methods reveals a well-articulated structure underlying the 20,675 IoT Interactions. MDS and LLE fail at recovering clear structure. PCA, Isomap and Spectral Embedding all reveal three groups of interactions that have relatively dissimilar if- and then-phrases (e.g., three somewhat separated orange clouds can be noted), but little else can be discerned. Of the six, t-SNE provides the greatest evidence of structure, though this structure is only apparent at the local level, without clear insight into higher level groupings. This is not surprising, given that t-SNE, a manifold learning algorithm for nonlinear dimensionality reduction (van der Maaten and Hinton 2008), preserves local similarity structure. Hence even the t-SNE results do not go far enough to generate a clear understanding of the structure of functional cosine similarity. The t-SNE results are suggestive of grouping tendencies, but the groups are not clearly delineated nor structured.

TDA Mapper uses the visualizations in Figure B1 as an intermediate result upon which a higher-order structure can be imposed. Based on these different lenses, the final topological models from TDA Mapper are shown in Figure B2. Only the t-SNE lenses produced a clearly interpretable topological model with a clear articulation of structure. This is due to the t-SNE's superior ability to represent local structure. Therefore, we used the two t-SNE coordinates as lenses for the TDA. The topological model shown in Figure 1 of the paper is based on the t-SNE lenses. A topological model can be interpreted by coloring its nodes according to characteristics of interest. For example, Figure 1 and Figure B2 color the nodes by average within similarity of the interactions in a node.

**Figure B1.** Dimensionality Reduction of 20,675 IFTTT Interactions Using Six Unsupervised Machine Learning Methods.

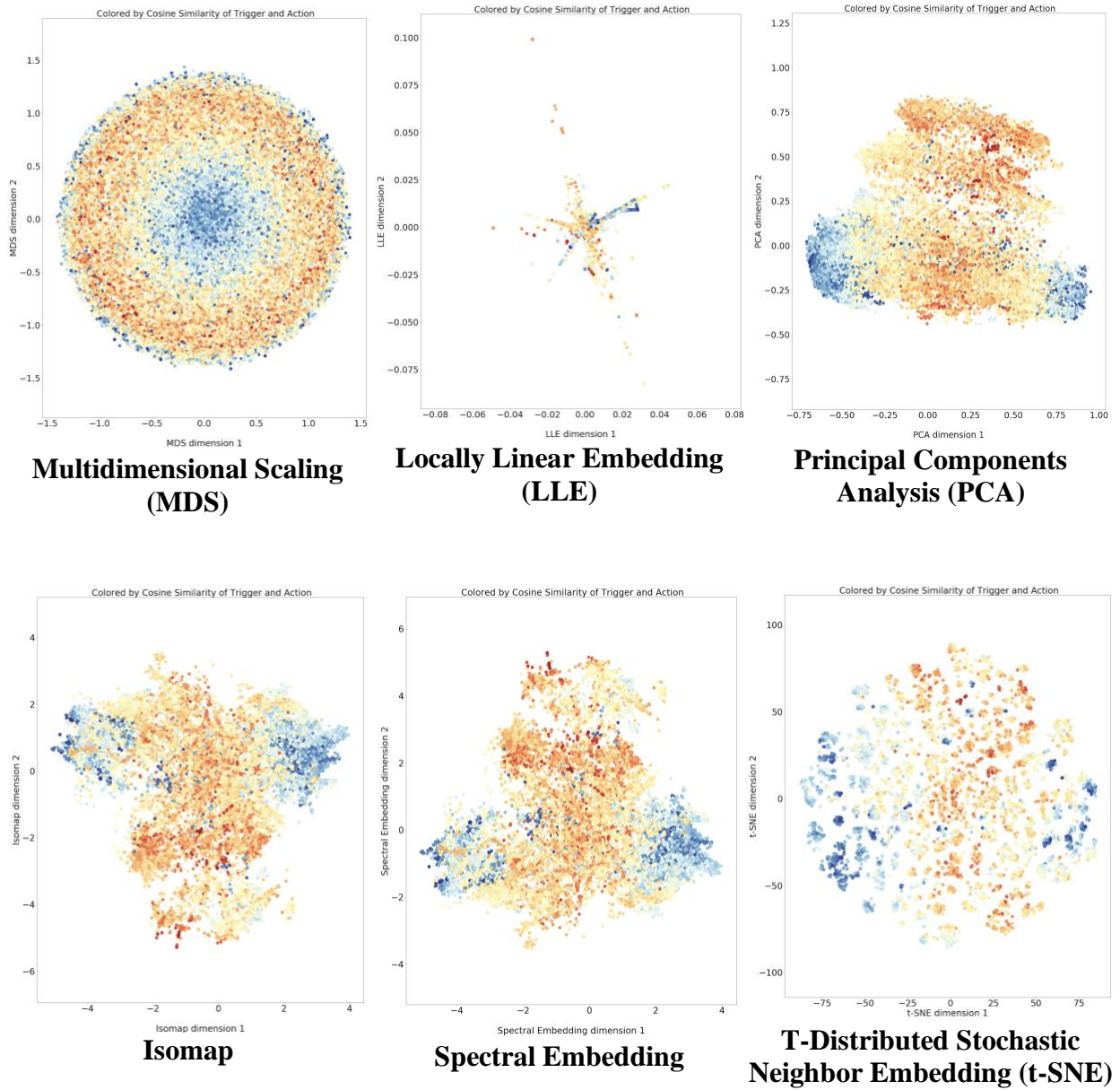

Note: Color indicates the degree of functional cosine similarity of each interaction's if- and then-embeddings, with blue indicating high cosine similarity (close to 1) and red indicating low cosine similarity (close to 0).

**Figure B2.** Topological Models Based Upon Two Dimensional Lenses from Six Unsupervised Machine Learning Methods.

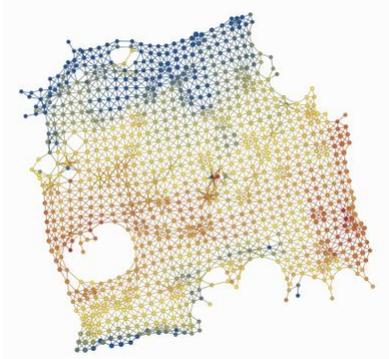
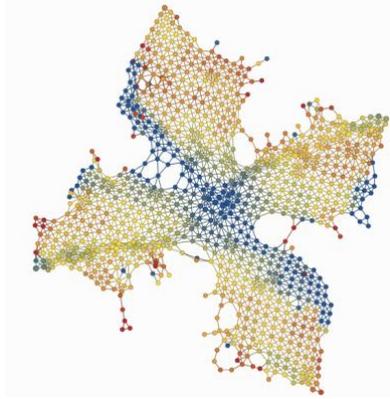
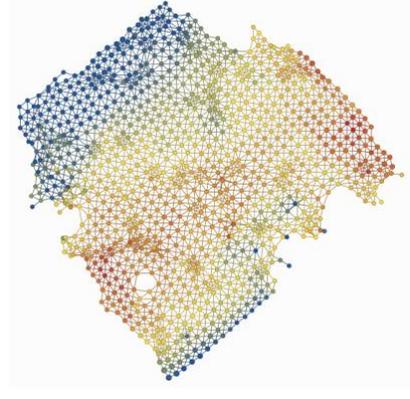

**Multidimensional Scaling (MDS) Lens**  **Locally Linear Embedding (LLE) Lens**  **Principal Components Analysis (PCA) Lens**

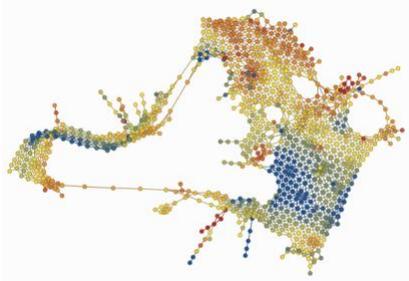
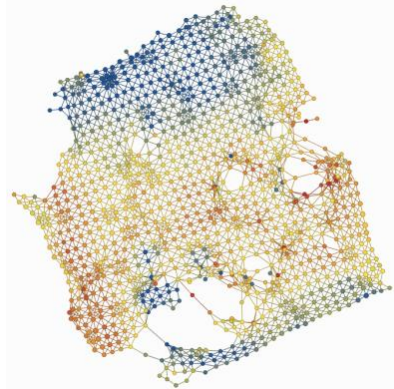
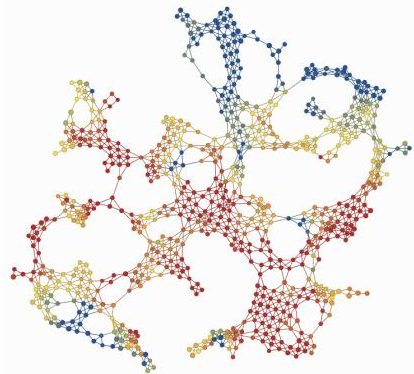

**Isomap Lens**  **Spectral Embedding Lens**  **T-Distributed Stochastic Neighbor Embedding (t-SNE) Lens**

Note: Color indicates the degree of functional cosine similarity of each interaction's if- and then-embeddings, with blue indicating high cosine similarity (close to 1) and red indicating low cosine similarity (close to 0).

## Opportunities for Exploration and Exploration by Louvain Community Groups

The topological model shown in Figure 1 of the paper reveals there are many different types of IoT interactions. It is possible that opportunities for exploration and exploration, as revealed by $r_{FUNCTIONAL, CC}$, may differ by type of IoT interaction. To identify these types of interactions, we used the Louvain modularity optimization algorithm for community detection (Blondel, et.al. 2008) to identify 18 groups of nodes in the topological model. The Louvain method is a network clustering algorithm that detects communities of similar interactions by optimizing modularity. The 18 community groups identified by the algorithm are summarized in Table B1 and overlaid on Figure B3 (this corresponds to Figure 1 in the paper). The numbering of groups begins with zero due to Python's 0-based indexing and numbers are not sequential due to the ten small "singleton" groups comprising 53 interactions that were excluded from the topological model. Table B1 shows a typical example of interactions in each community group.

The first column of Table B1 organizes the 18 community groups into six higher-order categories. We identified the six categories by combining adjacent community groups in the topological model, based on our expert judgment of the substantive meaning of each group. The six categories reflect the broader types of IoT automation experiences consumers have when connecting smart devices and include automating the smart home, smart home relationships, managing projects and content, communicating social content, saving new content for later, and being notified in real time of content. These six categories are shown in Figure 1 of the paper to aid interpretation of the topological model.

Finally, Figure B4 plots, within each of the 18 community groups, the relationship of within similarity of a focal interaction (horizontal axis) to the correlation of $cos_{FUNCTIONAL}$ and $cos_{CC}$ (vertical axis) shown in Figure 3 of the paper. As noted in the paper, incorporating an "interaction type" factor (operationalized by the 18 community groups) significantly improves the fit of a piecewise regression model. Figure B4 shows that there are clear and distinct differences among community groups in the within similarity of interactions along the horizontal axis. Correspondingly, the community groups differ in their potential for differentiating exploitation opportunities from exploration opportunities, as reflected by the correlations of functional and cross-capacity similarities along the vertical axis.

**Table B1.** 18 Community Groups Identified in Topological Model using the Louvain Modularity Optimization Algorithm for Network Clustering

| Higher-Order Category | Community Group # | Size (n) | Typical Interaction in Community Group |
|---|---|---|---|
| **Smart Home Automation** | 0 | 1079 | "If calendar event starts, set zone temperature" |
| | 1 | 394 | "If you enter or exit area, blink a light" |
| | 5 | 1209 | "If temperature rises above threshold, turn off A/C" |
| | 13 | 2051 | "If new motion after quiet period, blink lights" |
| | 14 | 1058 | "If you exit area, turn off heat" |
| **Smart Home Relationship** | 6 | 1052 | "If sleep duration above threshold, post Facebook update" |
| | 23 | 875 | "If soil moisture alert, update phone wallpaper" |
| | 24 | 1493 | "If nest protect battery low, send notification" |
| **Project and Content Management** | 18 | 716 | "If new trello card added, create google calendar event" |
| | 21 | 1805 | "If new iOS contact, append google drive document" |
| **Social Communication** | 4 | 1604 | "If new instagram photo of you, share update on linkedin" |
| | 9 | 968 | "If new task completed, post tweet" |
| | 10 | 1982 | "If new starred email, create facebook status message" |
| **Content Archiving** | 7 | 1474 | "If new feed item matches, save for later on pocket" |
| | 11 | 1027 | "If new instapaper archived item, create onenote" |
| | 15 | 610 | "If new wordpress post, add public bookmark" |
| **IOT Notifications** | 2 | 1007 | "If new closed issue on github, add to weekly email digest" |
| | 16 | 1552 | "If new NYT article, text me" |

**Figure B3.** 18 Community Groups in the Topological Model of 20,675 Interaction (Corresponds to Figure 1 of the Paper).

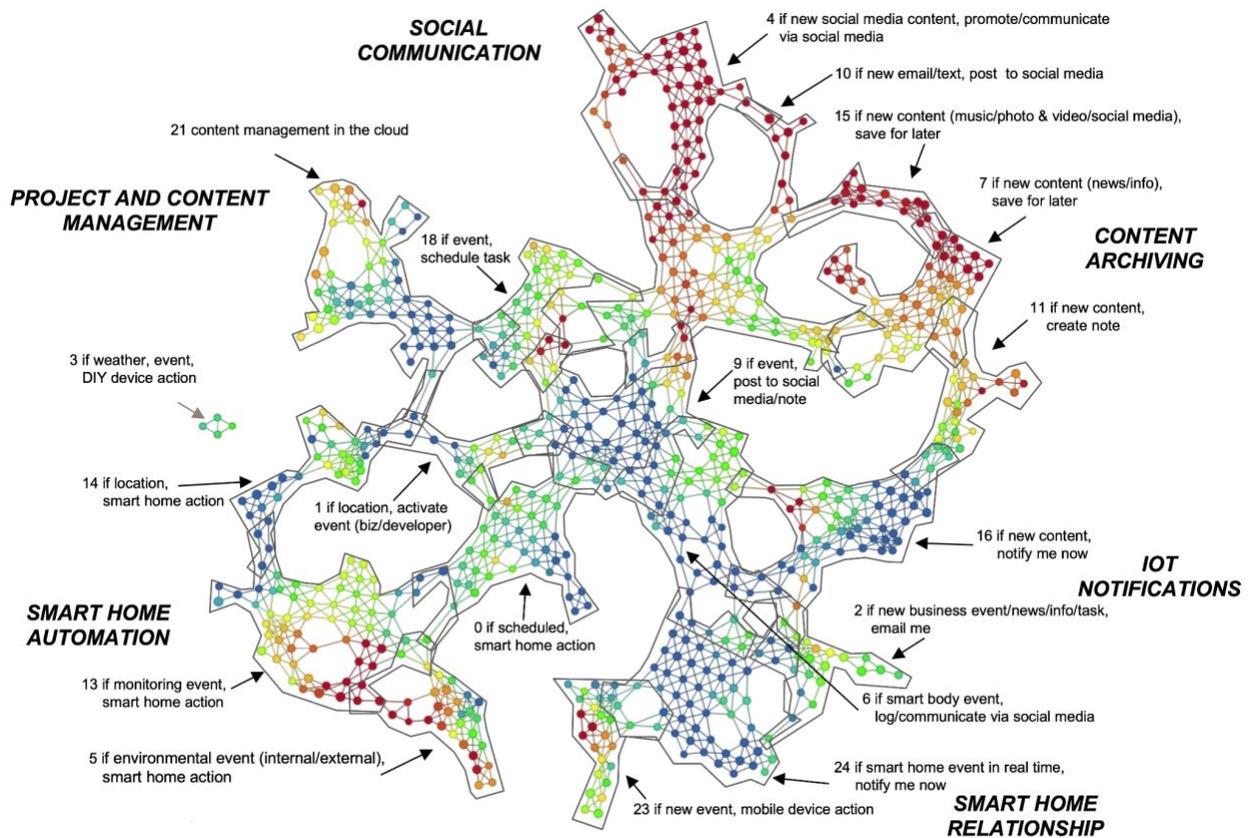

Network is colored by average cos<sub>WITHIN</sub> of the interactions in each node(blue=low, red=high)

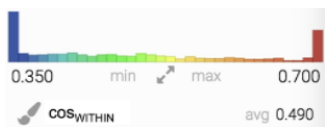

*Hyperparameters*

Lens: t-SNE
Distance metric: cosine distance
Resolution (number of bins): 35
Gain (degree of overlap): 2

**Figure B4.** Community Group Plots of Exploitation and Exploration.

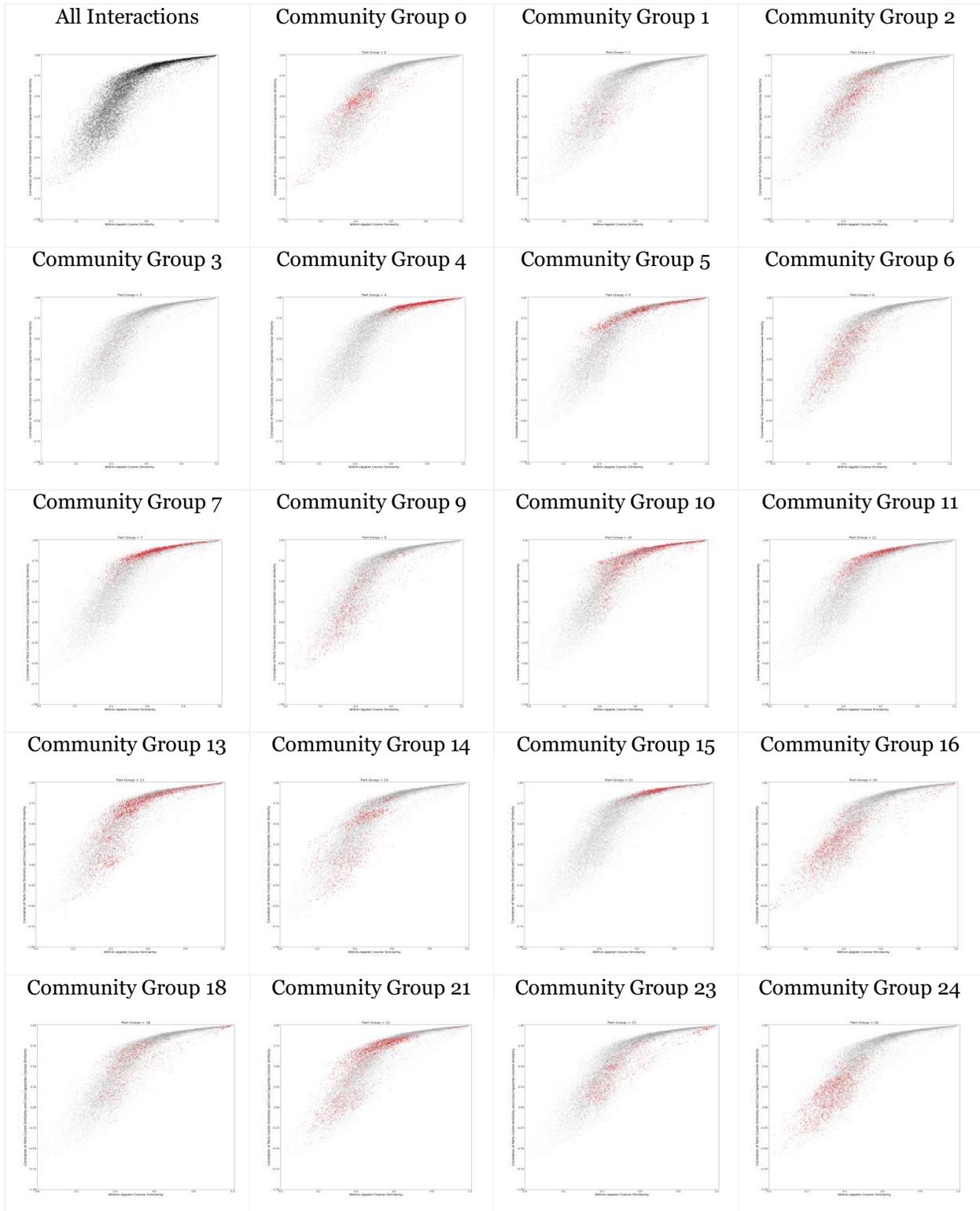

# Web Appendix C. Multiple Regression Predicting Correlation of Functional and Cross-Capacity Similarity from Within Similarity.

We further investigated the relationship between the within similarity of an interaction ($\cos_{WITHIN}$) and the correlation of functional and cross-capacity similarity of that interaction with other interactions ($r_{FUNCTIONAL, CC}$) by fitting a series of three nested regression models. The dependent variable in these three models is the correlation ($r_{FUNCTIONAL, CC}$) of functional ($\cos_{FUNCTIONAL}$) and cross-capacity ($\cos_{CC}$) similarity of each focal interaction with all other interactions. The independent variables are combinations of a) within similarity ($\cos_{WITHIN}$) of the focal interaction, b) a binary variable low defined as 1 if $\cos_{WITHIN}$ is below the mean (.509) and is otherwise 0, c) a categorical variable type containing community groups from Louvain modularity network clustering (described in Appendix B), and d) various interactions of these three independent variables. Tables C1 through C3 show, respectively, the omnibus F tests for Models 1 through 3 along with plots of predicted values from each model. Table C4 summarizes model fit measures for Models 1 through 3, and Table 5 provides tests of model comparisons for Models 1 versus 2, and Models 2 versus 3.

In Model 1, we predicted $r_{FUNCTIONAL, CC}$ from $\cos_{WITHIN}$ using a piecewise regression model. Based on Figure 3 in the paper, we fit separate intercept and slopes below and above the mean of $\cos_{WITHIN}$ (i.e., for $\cos_{WITHIN} < .509$ and for $\cos_{WITHIN} \geq .509$). Model 1 ($R^2 = .815$, AIC = -19183) fit better than a using a single slope and intercept ($R^2 = .736$, AIC = -11889), and also fit better than a non-piecewise model using linear, quadratic, and cubic terms for $\cos_{WITHIN}$ ($R^2 = .813$, AIC = -19054). Given the large sample size (n=20,657), the p-value of all F tests was <.001, and all $R^2$ and $R^2_{adjusted}$ values were identical.

We then aimed to determine how much of the variation in $r_{FUNCTIONAL, CC}$ (vertical axis) at values of $\cos_{WITHIN}$ (horizontal axis), evident in Figure 3, could be due to differences among types of interactions. To identify types of interactions, we used the Louvain modularity optimization algorithm for community detection (Blondel, et.al. 2008), a method for network clustering, to identify 18 groups of nodes in the topological model shown in Figure 1 of the paper. An interaction type factor, based on these groups of nodes, was included in Model 2 as a categorical predictor, along with the piecewise intercept and slopes from Model 1. Model 2 ($R^2$ = .848, AIC = -23224) has a superior fit than Model 1.

Lastly, we examined if the interaction type factor had different effects below and above the mean of $\cos_{\text{WITHIN}}$. Therefore, Model 3 incorporated additional terms beyond those in Model 2, allowing us to estimate piecewise effects of the interaction type factor. We found that Model 3 ($R^2$ = .862, AIC = -25172) has a superior fit than Model 2.

These preliminary results suggest three important implications. First, Model 1 implies that the ability to find opportunities for both exploitation and exploration increases as the similarity of the if- and then- parts of an interaction ($\cos_{\text{WITHIN}}$) decreases, particularly when within similarity is below average. Second, Model 2 indicates that, additionally, the ability to find opportunities for exploration varies across different types of interactions. Third, Model 3 demonstrates that the effects of different types of interactions on the opportunities for exploitation and exploration are also most impactful when within similarity is below average.

**Table C1.** Piecewise Regression Predicting $r_{FUNCTIONAL,\ CC}$ from $\cos_{WITHIN}$

## Model 1 - Omnibus ANOVA Test

|  | Sum of Squares | df | Mean Square | F | p |
|---|---|---|---|---|---|
| $\cos_{WITHIN}$ | 75.70 | 1 | 75.6973 | 3271 | < .001 |
| low | 7.85 | 1 | 7.8475 | 339 | < .001 |
| $\cos_{WITHIN}$ * low | 187.67 | 1 | 187.6749 | 8109 | < .001 |
| Residuals | 478.42 | 20671 | 0.0231 |  |  |

*Notes:*
1) Type 3 sum of squares
2) low = 1 if $\cos_{WITHIN}$ < .509, low = 0 if $\cos_{WITHIN}$ ≥ .509

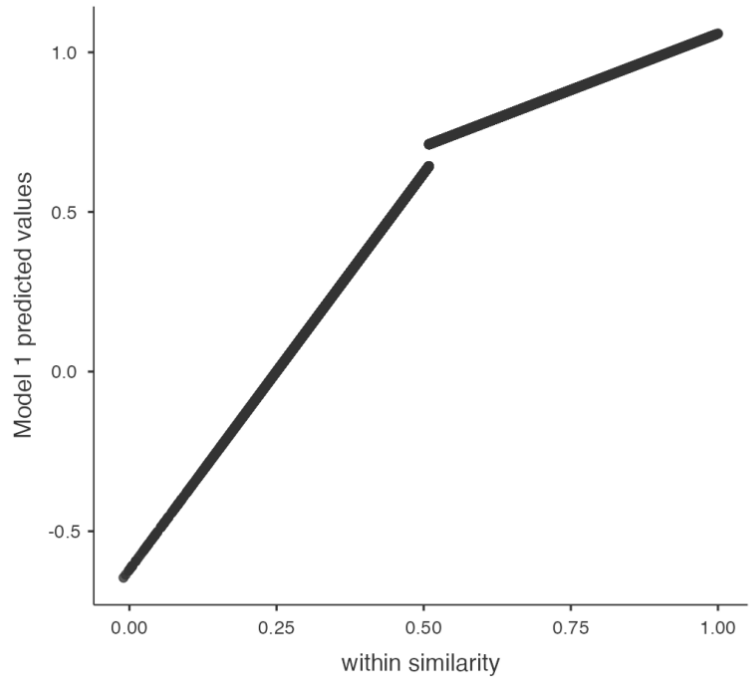

**Table C2.** Piecewise Regression Predicting $r_{FUNCTIONAL, CC}$ from $cos_{WITHIN}$ with Separate Intercepts for 18 Louvain Community Groups.

**Model 2 - Omnibus ANOVA Test**

|  | Sum of Squares | df | Mean Square | F | p |
|---|---|---|---|---|---|
| $cos_{WITHIN}$ | 51.42 | 1 | 51.41 | 51.4223 | 2704 |
| low | 2.21 | 1 | 2.2053 | 116 | <.001 |
| $cos_{WITHIN}$ * low | 139.10 | 1 | 139.0994 | 7315 | <.001 |
| type | 85.71 | 20 | 4.2853 | 225 | <.001 |
| Residuals | 3 | 20671 | 0.0190 | | |

*Notes:*
1) Type 3 sum of squares
2) low = 1 if $cos_{WITHIN}$ < .509, low = 0 if $cos_{WITHIN}$ ≥ .509
3) type = 18 Louvain community groups

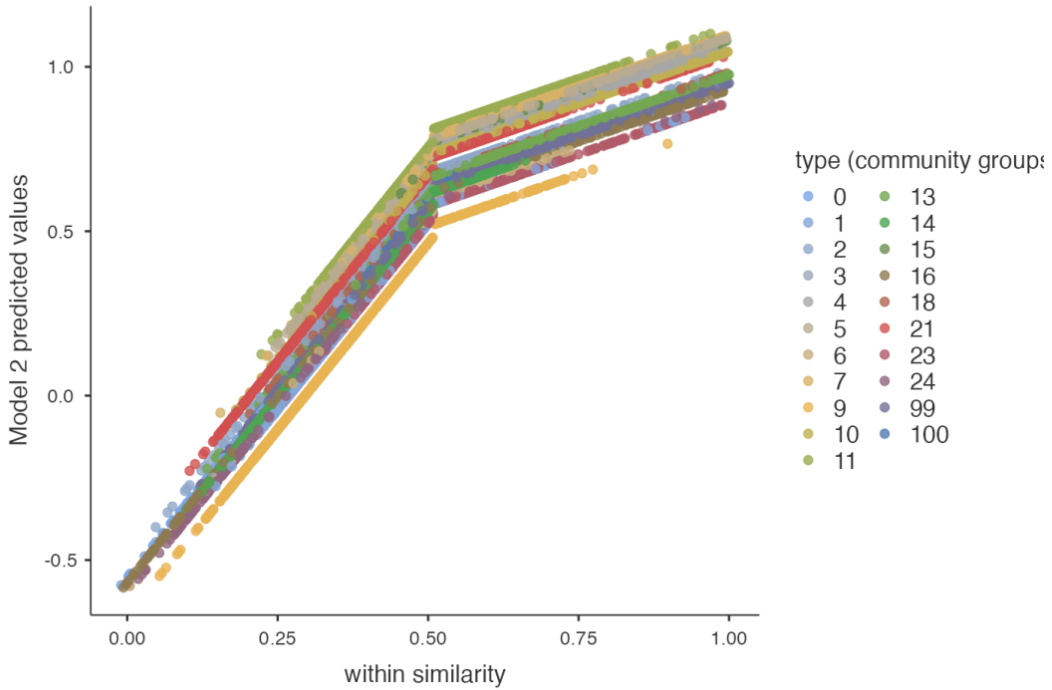

**Predicted Values from Model 2**

**Table C3.** Piecewise Regression Predicting $r_{FUNCTIONAL, CC}$ from $cos_{WITHIN}$ with Separate Intercepts and Slopes for 18 Louvain Community Groups.

### Model 3 - Omnibus ANOVA Test

|  | Sum of Squares | df | Mean Square | F | p |
|---:|---:|---:|---:|---:|---:|
| $cos_{WITHIN}$ | 1.458 | 1 | 1.4580 | 84.49 | < .001 |
| low | 0.990 | 1 | 0.9896 | 57.34 | < .001 |
| $cos_{WITHIN}$ * low | 1.460 | 1 | 1.4595 | 84.58 | < .001 |
| type | 24.432 | 20 | 1.2216 | 70.79 | < .001 |
| $cos_{WITHIN}$ * type | 4.569 | 20 | 0.2284 | 13.24 | < .001 |
| low * type | 2.536 | 20 | 0.1268 | 7.35 | < .001 |
| $cos_{WITHIN}$ * low * type | 9.604 | 20 | 0.4802 | 27.83 | < .001 |
| Residuals | 355.337 | 20591 | 0.0173 |  |  |

*Notes:*
1) Type 3 sum of squares
2) low = 1 if $cos_{WITHIN}$ < .509, low = 0 if $cos_{WITHIN}$ ≥ .509
3) type = 18 Louvain community groups

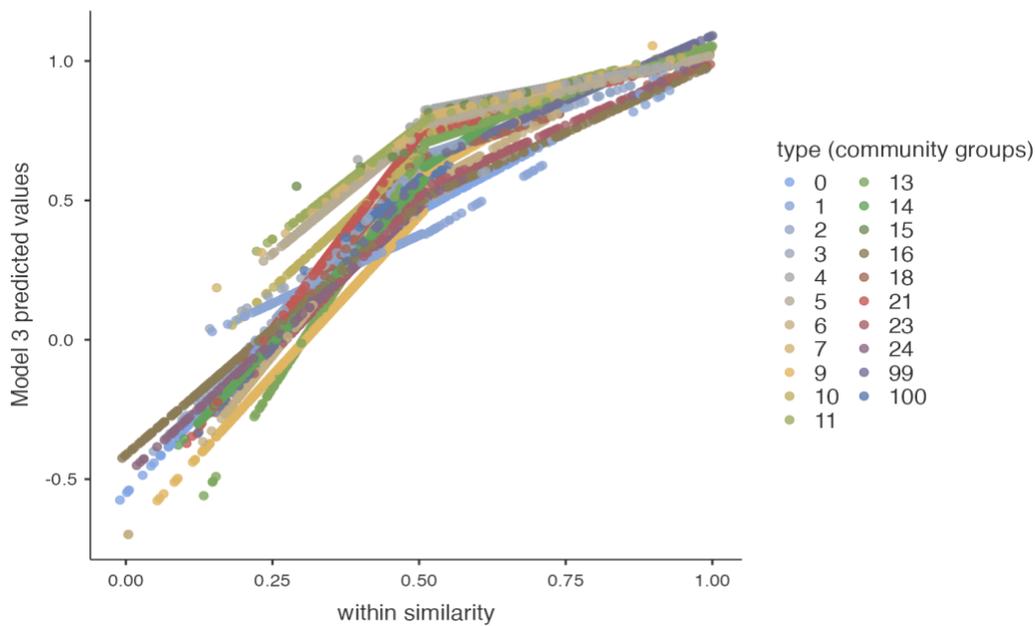

**Predicted Values from Model 3**

**Table C4.** Model Fit Measures.

| Model | R² | Adjusted R² | AIC | BIC |
|---|---|---|---|---|
| 1 | 0.815 | 0.815 | -19183 | -19143 |
| 2 | 0.848 | 0.848 | -23224 | -23026 |
| 3 | 0.862 | 0.862 | -25172 | -24497 |

**Table C5.** Model Comparisons.

| Comparison | ΔR² | F | df1 | df2 | p |
|---|---|---|---|---|---|
| Model 1 - Model 2 | 0.0332 | 225.3 | 20 | 20651 | < .001 |
| Model 2 - Model 3 | 0.0145 | 36.1 | 60 | 20591 | < .001 |

# WEB APPENDIX REFERENCES

Ayasdi (2018), "Ayasdi Machine Intelligence Platform, V7.12.0," Ayasdi, Inc. Menlo Park, CA

Blondel, Vincent D, Jean-Loup Guillaume, Renaud Lambiotte, and Etienne Lefebvre (2008), "Fast Unfolding of Communities in Large Networks," *Journal of Statistical Mechanics: Theory and Experiment*, 10, October, 10008.

Mikolov, Tomas, Kai Chen, Greg Corrado, and Jeffrey Dean (2013a), "Efficient Estimation of Word Representations in Vector Space," CoRR , abs/1301.3781.

Novak, Thomas P., and Donna L. Hoffman (2023), "Automation Assemblages in the Internet of Things: Discovering Qualitative Practices at the Boundaries of Quantitative Change," *Journal of Consumer Research,* 49 (5), 811-837.

Ray, Jeffrey and Marcello Trovati (2017), "A Survey of Topological Data Analysis (TDA) Methods Implemented in Python," in *International Conference on Intelligent Networking and Collaborative Systems*, 594-600. Springer, Cham.

Singh, Gurjeet, Facundo Mémoli, and Gunnar E. Carlsson. (2007), "Topological Methods for the Analysis of High Dimensional Data Sets and 3d Object Recognition," In *SPBG*, pp. 91-100. Soulier et. al 2012

van der Maaten, Laurens, and Geoffrey Hinton (2008), "Visualizing Data Using t-SNE," *Journal of Machine Learning Research*, 9 (Nov), 2579-2605.